\documentclass[journal=jacsat,manuscript=article]{achemso}
\setkeys{acs}{etalmode=truncate,maxauthors=0}

\usepackage{chemformula} % Formula subscripts using \ch{}
\usepackage[T1]{fontenc} % Use modern font encodings
\usepackage{graphicx}
\usepackage{natbib}
\usepackage{color,bm,braket}
\setkeys{acs}{keywords = true}

\newcommand {\mo}{MoS$_2$}
\newcommand {\mose}{MoSe$_2$}
\newcommand {\lm}{\lambda_{\rm MTB}}
\newcommand {\lmb}{\lambda_{\rm MTB}^{\rm band}}
\newcommand {\lmp}{\lambda_{\rm MTB}^{\rm pol}}

\author{Clifford Murray}
\email{murray@ph2.uni-koeln.de}
\author{Camiel van Efferen}
\author{Wouter Jolie}
\affiliation[Uni Koeln]
{II. Physikalisches Institut, Universit\"{a}t zu K\"{o}ln, Cologne D-50937, Germany}
\alsoaffiliation[Uni Muenster]
{Institut f{\"ur} Materialphysik, Westf\"{a}lische Wilhelms-Universit\"{a}t M\"{u}nster, M\"{u}nster D-48149, Germany}
\author{Jeison Antonio Fischer}
\author{Joshua Hall}
\affiliation[Uni Koeln]
{II. Physikalisches Institut, Universit\"{a}t zu K\"{o}ln, Cologne D-50937, Germany}
\author{Achim Rosch}
\affiliation[THP]
{Institut f\"{u}r Theoretische Physik, Universit\"{a}t zu K\"{o}ln, Cologne D-50937, Germany}
\author{Arkady V. Krasheninnikov}
\affiliation[Dresden]
{Institute of Ion Beam Physics and Materials Research, Helmholtz-Zentrum Dresden-Rossendorf, Dresden D-01328, Germany}
\alsoaffiliation[Aalto]
{Department of Applied Physics, Aalto University School of Science, Aalto FI-00076, Finland}
\author{Hannu-Pekka Komsa}
\affiliation[Aalto]
{Department of Applied Physics, Aalto University School of Science, Aalto FI-00076, Finland}
\alsoaffiliation[Oulu]
{Microelectronics Research Unit, University of Oulu, Oulu FI-90014, Finland}
\author{Thomas Michely}
\affiliation[Uni Koeln]
{II. Physikalisches Institut, Universit\"{a}t zu K\"{o}ln, Cologne D-50937, Germany}

\title{Band Bending and Valence Band Quantization at Line Defects in MoS$_2$}

\begin{document}

\begin{abstract}
  The variation of the electronic structure normal to 1D defects in quasi-freestanding MoS$_2$, grown by molecular beam epitaxy, is investigated through high resolution scanning tunneling spectroscopy at $5$\,K. Strong upwards bending of valence and conduction bands towards the line defects is found for the 4|4E mirror twin boundary and island edges, but not for the 4|4P mirror twin boundary. Quantized energy levels in the valence band are observed wherever upwards band bending takes place. Focusing on the common 4|4E mirror twin boundary, density functional theory calculations give an estimate of its charging, which agrees well with electrostatic modeling. We show that the line charge can also be assessed from the filling of the boundary-localized electronic band, whereby we provide a measurement of the theoretically predicted quantized polarization charge at MoS$_2$ mirror twin boundaries. These calculations elucidate the origin of band bending and charging at these 1D defects in MoS$_2$. The 4|4E mirror twin boundary not only impairs charge transport of electrons and holes due to band bending, but holes are additionally subject to a potential barrier, which is inferred from the independence of the quantized energy landscape on either side of the boundary.  
\end{abstract}

\bigskip

Coupled to the rise of MoS$_2$ and other transition metal dichalcogenide (TMDC) semiconductors as prospective two-dimensional (2D) device materials came the need to investigate their one-dimensional (1D) defect structures, \textit{e.g.} grain boundaries (GBs). Depending on their structure, GBs impair device performance to differing degrees when positioned in the channel of a single layer MoS$_2$ field effect transistor \cite{VanDerZande2013,Najmaei2014,Du2016,Ly2016}. It is thus evident that control of the type and concentration of GBs is of importance for device fabrication. Besides satisfying scientific curiosity, it therefore pays to understand their effect on band structure and charge carrier transport. The lowest energy GBs are those hardest to avoid during growth, as the energy penalty associated with their introduction is marginal. In the three-dimensional (3D) world, these low energy GBs are 2D stacking faults or twin planes. For the case of SiC devices such defects cause increased leakage current, reduced blocking voltage, and the degradation of bipolar devices \cite{Kimoto2015,Skowronski2006}. In the world of 2D materials, the analog to twin planes is 1D mirror twin boundaries (MTBs). These structural defects have some surprising effects on the band structure of monolayer \mo, to be investigated in this manuscript.

GBs in MoS$_2$ layers have already been intensively investigated. Experimentally, scanning tunneling microscopy (STM) and spectroscopy (STS) are the ideal tools to identify and characterize \mo~GBs electronically\cite{Huang2015,Wang2018b,Yan2018,Liu2016,Precner2018}. Despite large variations in magnitude, typically an upwards shift of the valence band (VB) and often also of the conduction band (CB) is observed at the GB \cite{Huang2015,Zhang2014,Wang2018b,Liu2016,Precner2018,Yan2018}. The magnitude of the reported shifts range from 0.15\,eV to 1.0\,eV for the VB, depending on substrate and GB type. CB shifts may be of the same magnitude, but are usually smaller. The spatial extension of the band bending is on the order of $5$\,nm away from the GB. Similar band bending effects were also described for \mo~\cite{Zhang2014}, MoSe$_2$ and WSe$_2$ edges \cite{Zhang2016,LeQuang2018}, as well as for lateral heterojunctions, \textit{e.g.} of MoS$_2$ with WS$_2$ or WSe$_2$ \cite{Kobayashi2016,Zhang2018}. As an explanation for the observed band bending strain \cite{Huang2015,Wang2018b,Ma2017b,Zhang2018}, charge transfer into in-gap states \cite{Zhang2014,Ly2016,Kaneko2018}, or a combination of both were invoked \cite{Liu2016,Yan2018,Kobayashi2016}. Despite the technological importance of the band bending, modifications of the electronic structure around GBs are still poorly understood.

Here, for the investigation of 1D line defects in MoS$_2$, we take an approach different from previous work. Molecular beam epitaxy (MBE)\cite{Hall2017}, rather than the typical chemical vapor deposition, is employed for the growth of the MoS$_2$ layer to be investigated by STM and STS. The advantage is that the MoS$_2$ layer is grown under ultra-high vacuum conditions and remains thereunder for spectroscopic investigation. Thereby the MoS$_2$ and the potentially reactive 1D defects remain clean of adsorbates. Inert, single-crystal graphene (Gr) on Ir(111) is used as a substrate \cite{Coraux2008}. Consequently, high-resolution STS is facilitated and ambiguities in the comparison of the data to density functional theory (DFT) calculation, which could result from adsorption or an inhomogeneous environment, are avoided. We discover that band bending over a distance of $5$\,nm and the associated confinement normal to the 1D line defect renders the \mo~VB quantized. Although we focus our investigation on the most frequent and presumably lowest energy MTBs \cite{Komsa2017,Jolie2019}, we find the same phenomenon for the edges of monolayer (ML) and bilayer (BL) \mo~islands, as well as for less symmetric, higher energy GBs. From the comparison between the experimental and calculated results, we establish that the band bending is not caused by strain but by charge on the line defects. Using DFT and measuring the Fermi wave vectors of MTB states, we decompose the net charge on the line defects into contributions from polarization and in-gap states of the 1D line defects. Charge transfer from Gr into in-gap defect bands is of decisive importance for band bending and the formation of the hole confining potential in the VB, meaning that this effect is substrate-tunable.
 
\section{Results and Discussion}
\begin{figure}
	\centering
	\includegraphics[width=160mm]{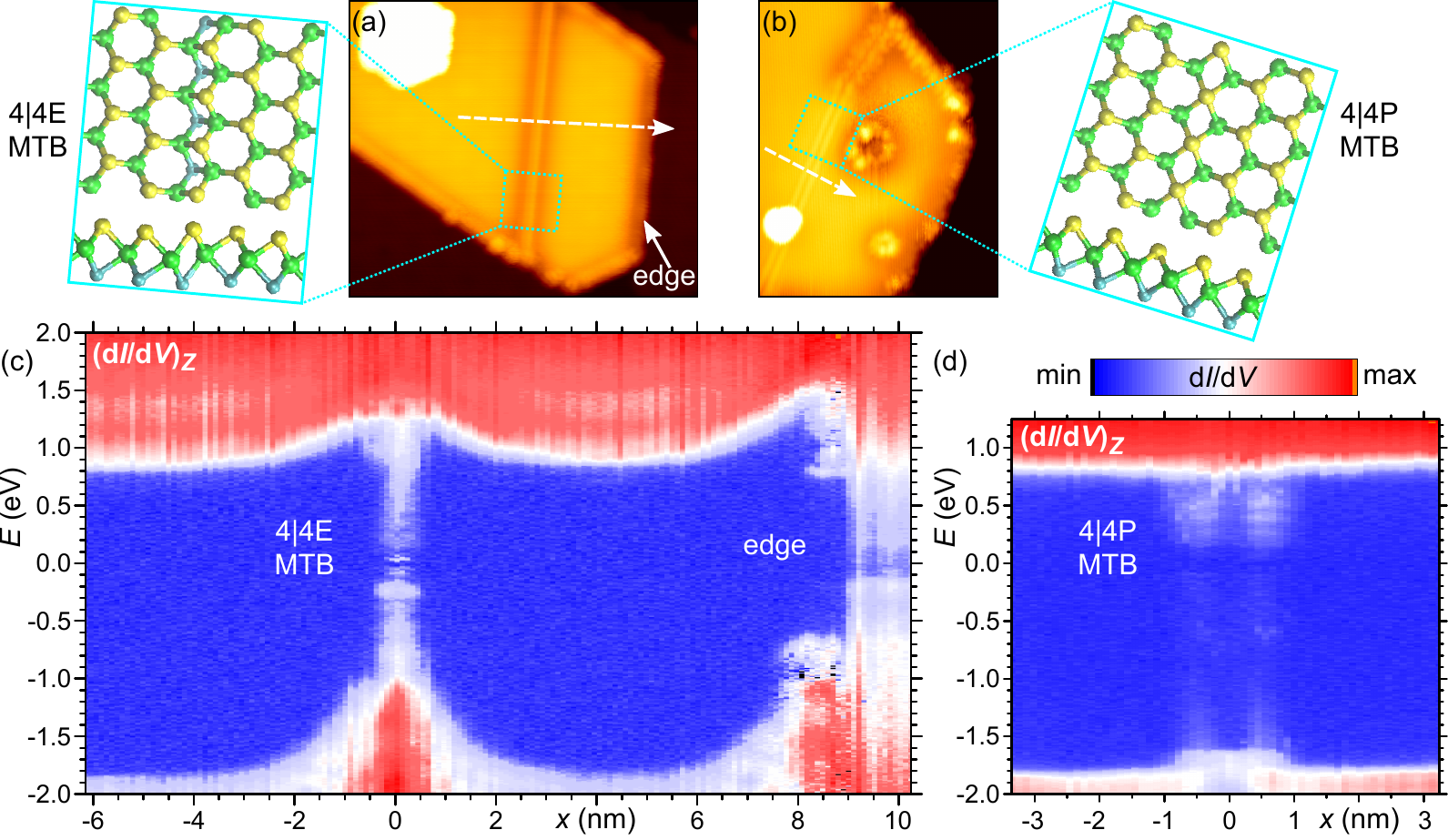}
	\caption{STS linescans across 1D defects in ML-\mo~on Gr/Ir(111). (a,b) Constant current STM topographs showing ML-\mo~islands; Gr appears as a dark background. In (a) a $\approx20$\,nm long 4|4E-type MTB runs near-vertically down the center of the image, with a straight island edge parallel to it. In the top-left of the image a small BL-\mo~island is seen. In (b) a 4|4P-type MTB is present. At the top of the image the MTB terminates and an irregular GB begins, while bottom-left a bright object (probably a Mo cluster) lies on top of the MTB. Ball-and-stick models of the 4|4E and 4|4P MTBs, in top and side views, are shown according to their topograph orientation in (a) and (b), respectively. Mo atoms are green, S atoms are yellow (top layer) or cyan (bottom layer). (c,d) Constant height STS linescans acquired along the dashed white arrows shown in (a) and (b) respectively. The recorded $(\text{d}I/\text{d}V)_Z$ signal is plotted as a function of energy $E$ and position $x$, according to the shown (logarithmic) color scale with arbitrary units. STM/STS parameters: (a) $V=1.2$\,V, $I=50$\,pA, image size $27\times23$\,nm$^\text{2}$; (b) $V=1.0$\,V, $I=100$\,pA, image size $19\times15$\,nm$^\text{2}$; (c) stabilization voltage $V_\text{st}=2.00$\,V, stabilization current $I_\text{st}=800$\,pA; (d) $V_\text{st}=1.25$\,V, $I_\text{st}=50$\,pA.
		\label{fig1}}
\end{figure}

\textbf{Scanning Tunneling Spectroscopy.}
The STM topographs in Figures~\ref{fig1}a and \ref{fig1}b show the 1D defects typical for MBE-grown ML-\mo~islands, in our case on Gr/Ir(111) (see Methods). A 4|4E- and a 4|4P-type MTB are visible in Figures~\ref{fig1}a and \ref{fig1}b, respectively; their atomic structures are depicted, illustrating that they form between $180^\circ$-misoriented domains. 4|4E- or 4|4P-type MTBs appear in topographs as bright single- or double-tracks, respectively (see Ref.~\citenum{Jolie2019} for details). The MTBs host 1D metallic states and consist of 4-fold rings sharing an \textbf{e}dge along a Mo-S bond (hence named 4|4E) or sharing a \textbf{p}oint at S-dimer sites (hence named 4|4P) \cite{Zhou2013,Zou2013,Zou2015,Jolie2019}. In \mose~and MoTe$_2$ the 4|4P-type MTB is energetically most favorable, while in \mo~both 4|4P- and 4|4E-types are stable \cite{Komsa2017,Batzill2018}. In our samples, 4|4E-type MTBs are about $5$ times more common and typically longer than 4|4P-type. An atomically straight zigzag island edge, parallel to the MTB, is also indicated in Figure~\ref{fig1}a.

We perform high-resolution, constant height STS linescans orthogonally across the 1D defects; in Figure~\ref{fig1}c over the 4|4E MTB and continuing over the island edge, in Figure~\ref{fig1}d over the 4|4P MTB. The meV-, \AA-resolution of the linescans leaves no data interpolation or filtering necessary (see Methods). In the colorplot, blue represents a low or zero signal intensity (\textit{i.e.} in the band gap) while the red end of the scale signifies a finite density of states. The unperturbed or `bulk' ML-\mo~band structure is measured when sufficiently far from defects, for example in Figure~\ref{fig1}c at $x=-6$\,nm: at $\approx0.8$\,eV the onset of the CB and at $\approx-1.8$\,eV that of the VB are seen by STS.\cite{Murray2019} 

Scanning orthogonally over the 4|4E MTB, defined to lie at $0$\,nm in Figure~\ref{fig1}c, both the VB and CB bend several hundred meV upwards (\textit{i.e.} towards higher energies). This bending occurs in a range of $\approx 5$\,nm on both sides of the MTB. No band bending of significance could be detected beyond this range \cite{Yu2016}.States are detected throughout the band gap at the location of the MTB. They are derived from the 1D metallic MTB bands in the band gap of the 2D MoS$_2$ \cite{Jolie2019}. Their quantization and Coulomb-blockade energy gap, associated with the Tomonaga-Luttinger liquid hosted along the finite-length MTB,\cite{Jolie2019} are visible close to $E_\text{F}$. Towards the island edge (located at $\approx8.5$\,nm) similar upwards bending and in-gap states are observed. Finally, the relatively featureless spectra of Gr on Ir(111) are recorded for $x>9$\,nm. In Figure~\ref{fig1}d the 4|4P-type MTB is located at $0$\,nm. The double-track structure of the 1D metallic states \cite{Jolie2019} is faintly visible. In contrast to the 4|4E MTBs and edges, the 4|4P MTB does not cause significant band bending. Further measurements confirming this are available in the Supporting Information, Figure~S1.

\begin{figure}
	\centering
	\includegraphics[width=65mm]{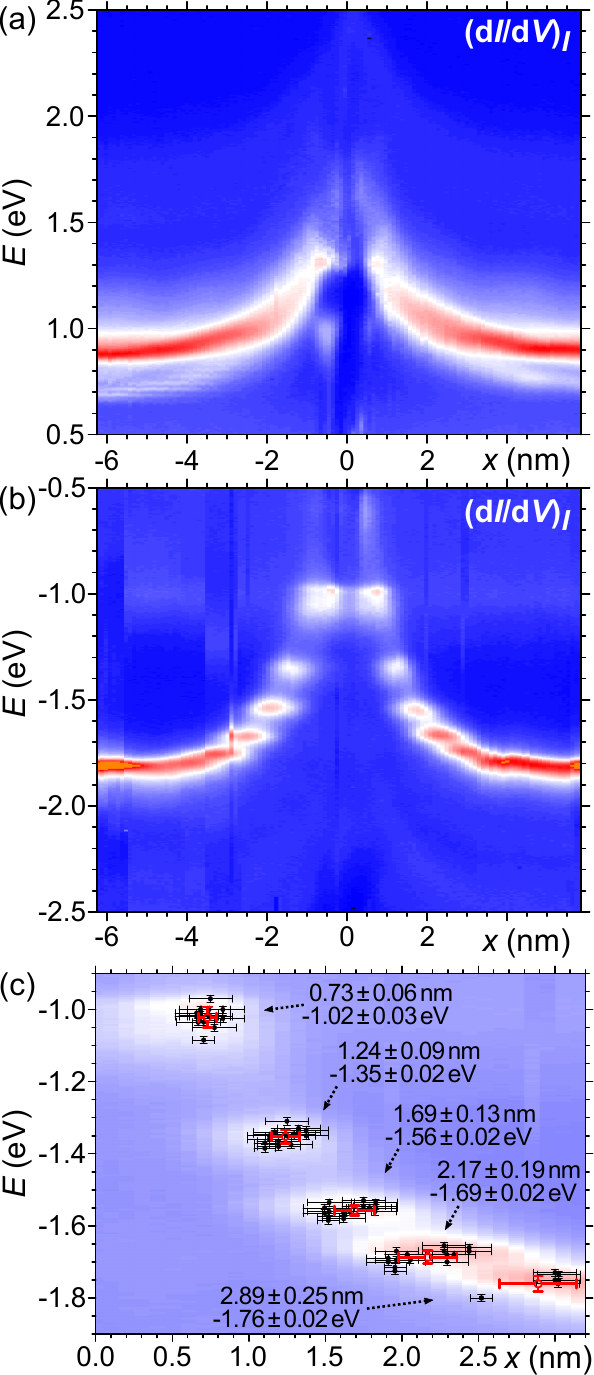}
	\caption{(a,b) Constant current STS linescans over a 4|4E MTB in ML-\mo~within the energy ranges of the CB and VB extrema, respectively. The recorded  $(\text{d}I/\text{d}V)_I$ signal is plotted as in Figure~\ref{fig1}d, but on a linear color scale. See Figure~S2 for a topograph of the MTB and a constant height STS scan in the same location. (c) Statistical distribution of quantized VB states at 4|4E MTBs. Black points show the energies $E$ and positions $x$ of the STS maxima from 20 different sets of states, mean values are in red and are stated. See Supporting Information for further details. For comparison, the right-hand side states of (b) are shown in the background. STS parameters: (a) $V_\text{st}=2.5$\,V; (b) $V_\text{st}=-2.5$\,V; (a,b) $I_\text{st}=50$\,pA.
		\label{fig2}}
\end{figure}

\clearpage
 We focus on the large band bending seen at 4|4E MTBs, and obtain additional information \textit{via} constant current STS; exemplary linescans are shown in Figures~\ref{fig2}a and \ref{fig2}b. Constant current STS is of higher sensitivity to the variations at the band edges, \textit{i.e.} to abrupt increases of the LDOS due to van Hove singularities, and thus can better resolve the subtle features at the edges, while features within the bands are hardly visible.\cite{Zhang2015,Stroscio1993a}. We note that similar band edge resolution with an inferior signal-to-noise ratio can be achieved by constant height STS after normalizing to the tunneling current (compare Figure S2). 

The 1D MTB band is metallic and was visible at $x=0$ in the constant height STS of Figure~\ref{fig1}c. However, with increasing magnitude of bias voltage no abrupt increase of the LDOS takes place in the investigated ranges of Figure~\ref{fig2}(a,b), and so the 1D MTB band is not visible in constant current STS. Well-visible in the constant current STS are the conduction and the valence band edges in Figure~\ref{fig2}a and Figure~\ref{fig2}b, respectively, through a sudden intensity rise.  

In the CB, Figure~\ref{fig2}a, far from the MTB we see the Q-point edge states at $\approx0.9$\,eV and, faintly, the K-point edge states at $\approx0.7$\,eV~\cite{Murray2019}. The latter is the CB minimum of ML-\mo~and is barely visible in constant height STS due to its large parallel momentum, but constant current STS allows the tip to move closer to the sample to detect this weak signal~\cite{Murray2019,Zhang2015}. Towards the MTB, the bands are bent smoothly up to roughly $1.3$\,eV. The K-point edge appears to bend slightly more than the Q-point edge, until their signals merge around $2$\,nm from the MTB. Bending of higher energy band edges is also faintly visible.

In the VB, Figure~\ref{fig2}b, the dominating $\Gamma$-point edge begins at around $-1.8$\,eV and bends upwards to $-1.0$\,eV at the MTB, thus undergoing larger bending than the CB. The VB bending occurs stepwise in increasing energy intervals. The discrete energies are near identical on either side of the MTB and were approximately the same across dozens of such MTBs of various lengths, see Figure~\ref{fig2}c. 

\begin{figure}
	\centering
	\includegraphics[width=160mm]{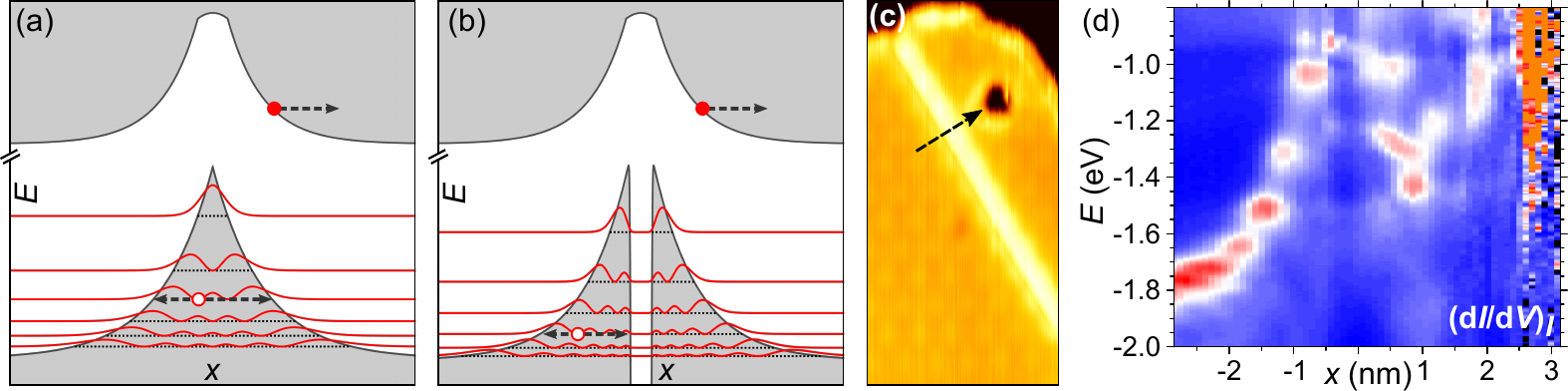}
	\caption{(a) Schematic sketch of hypothetical band bending at the 4|4E MTB, not to scale. Gray shaded areas represent the LDOS. Also shown are energy levels (black dotted lines) and probability densities (red, arbitrary scale) obtained by solving the 1D Schr\"{o}dinger equation for holes in this confining potential. (b) As (a), but assuming that the MTB gives rise to an additional repulsive potential, bisectioning the potential of (a) into two mirror-symmetric confining potentials. (c) Constant current STM topograph in which a small hole in the ML-\mo~island can be seen close to a 4|4E MTB. (d) Constant current STS linescan acquired along the black dashed arrow shown in (c). Linear color scale as Figure~\ref{fig2}.  STM/STS parameters: (c) $V=-2.0$\,V, $I=100$\,pA, image size $30\times15$\,nm$^\text{2}$; (d) $V_\text{st}=-2.0$\,V, $I_\text{st}=100$\,pA.
		\label{fig3}}
\end{figure}

We consider a simple explanation for these discrete energies: the bending of the VB creates a potential well as sketched in Figure~\ref{fig3}a. This confinement quantizes the VB energy levels perpendicular to the MTB, whereas electrons excited to the CB are not confined. The energy levels and the maxima of the squared wave functions at the edge of the potential are reasonably well reproduced by solving the 1D Schr\"{o}dinger equation for holes in the shown potential. The strongest probability density maxima of the quantized states are located close to the bent VB edge, with additional small peaks closer to the MTB. These smaller peaks are invisible to constant current STS (compare Figure~\ref{fig2}b), while constant height STS shows featureless intensity below the band edge next to the MTB (compare Figure~\ref{fig1}c). To understand this, it is important to note that the states represented in Figure~\ref{fig3}a are 1D solutions of the  Schr\"{o}dinger equation for a 1D potential \textit{normal} to the MTB. In reality they mark the van Hove singularities of confined states that disperse \textit{parallel} to the MTB. Their hole dispersion along the MTB is characterized by downwards, open parabolas which overlap throughout below the band edge, but not at the band edge. Thus the edge-sensitive constant current STS mode measures only the dominant van Hove singularities, while constant height STS measures a featureless LDOS below the bent VB edge due to overlapping states.

An alternative scenario to the single potential well consists of two independent wells, left and right of the MTB, with the MTB acting as a repulsive barrier. It allows one to recreate the observed peaks as well, as suggested by Figure~\ref{fig3}b. We note that distinction of these two situations is relevant, as hole transport through the MTB will be strongly impaired if the MTB acts as a repulsive potential for holes.

Experimentally it is straightforward to distinguish these two situations. Figure~\ref{fig3}c shows a \mo~island which features a small vacancy island directly next to a 4|4E MTB. An STS linescan from the complete side of the MTB to the broken side is displayed in Figure~\ref{fig3}d. On the complete side the quantized energy levels of the VB are like that at a `pristine' MTB, compare Figure~\ref{fig2}c. On the other side, however, the quantized states are strongly modified due to the changed potential energy landscape. The eigenenergies and eigenstates on the complete side are not perturbed by the environment on the broken side, making plain that the two sides are independent \textemdash~the MTB is a barrier to holes confirming the scenario of Figure~\ref{fig3}b.

\begin{figure}
	\centering
	\includegraphics[width=160mm]{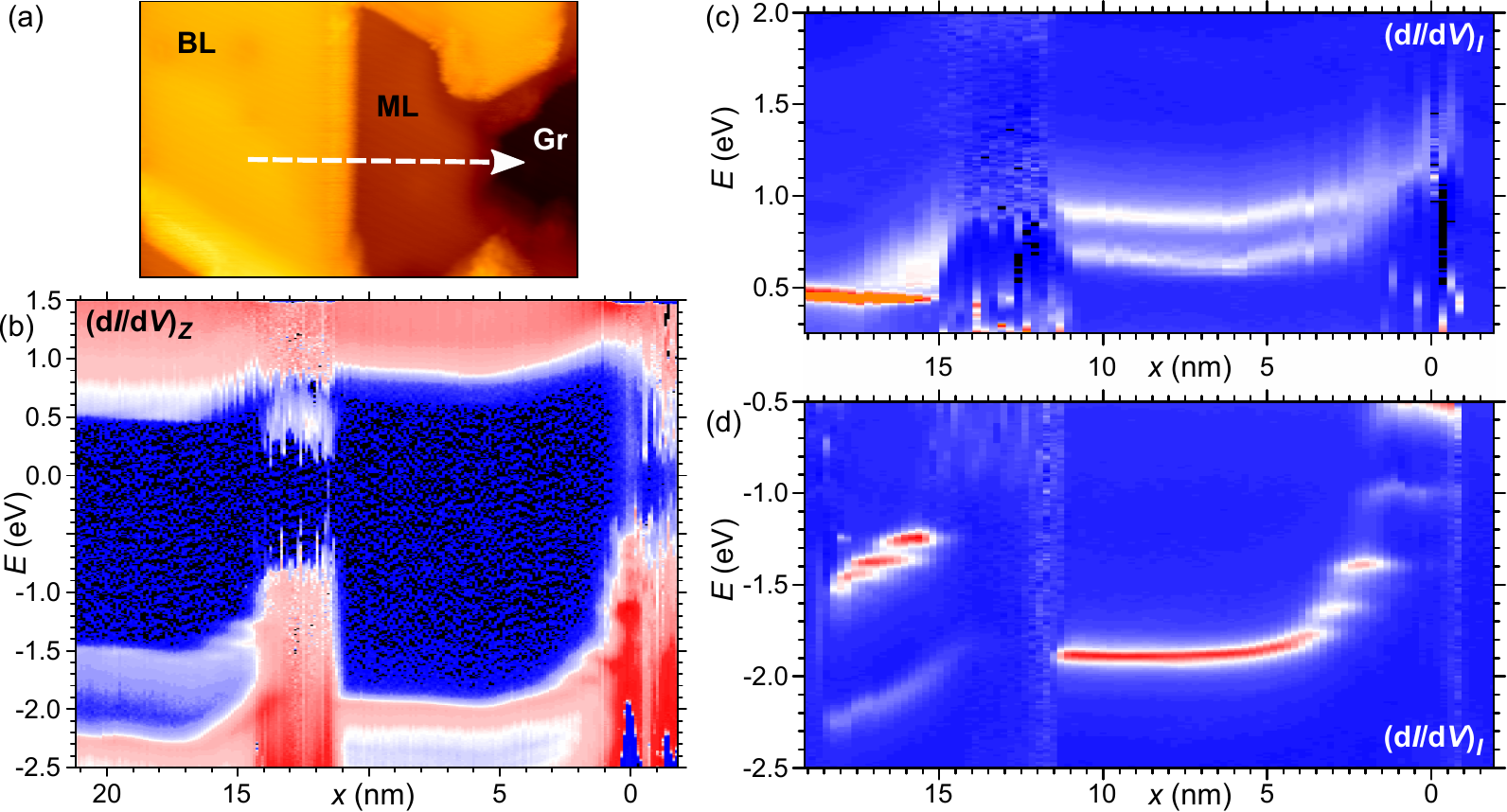}
	\caption{Band bending at edges of BL- and ML-\mo~islands. (a) Constant current STM topograph showing BL-\mo, ML-\mo, and Gr as indicated. The dashed white arrow marks where STS linescans were performed. (b) Constant height STS linescan, logarithmic color scale as Figure~\ref{fig1}d. (c,d) Constant current STS linescans of the conduction and valence band extrema, respectively. Linear color scales as Figure~\ref{fig2}. $x=0$\,nm defines the ML-\mo~edge in each scan. STM/STS parameters: (a) $V=1.0$\,V, $I=80$\,pA, image size $40\times25$\,nm$^\text{2}$; (b) $V_\text{st}=1.5$\,V, $I_\text{st}=90$\,pA; (c) $V_\text{st}=2.25$\,V; (d) $V_\text{st}=-2.50$\,V; (c,d) $I_\text{st}=80$\,pA.
		\label{fig4}}
\end{figure}

Does the VB bending at a \mo~island edge, as visible in Figure~\ref{fig1}c, cause quantization similar to that at a MTB? And does such band bending also occur at BL-\mo~island edges? To answer these questions we perform STS linescans along the path marked in Figure~\ref{fig4}a, from a BL-island \textit{via} BL- and ML-\mo~edges down to the Gr level. In constant height STS, Figure~\ref{fig4}b, it is seen that also the BL-\mo~bands undergo upwards bending at the island edge, similar to in the ML. Both of the BL's split $\Gamma$-point bands (at $\approx-1.4$\,eV and $\approx-2.2$\,eV at $x=20$\,nm) are bent upwards at the edge. More information is gathered from constant current STS of the CB and VB energy ranges shown in Figures~\ref{fig4}c and \ref{fig4}d, respectively. The BL CB edge appears to bend $\approx0.3$\,eV upwards, while in the ML bending of the K- and Q-point states (with slightly different curvatures) bears a resemblance to that at 4|4E MTBs (Figure~\ref{fig2}a) and is of the order of 0.5\,eV. Although STS at \mo~edges is typically affected by tip instabilities like those visible in Figure~\ref{fig4}d, quantization of the VB is unambiguous next to both the BL- and ML-\mo~island edges. While the discrete VB energy values were approximately the same for other ML island edges, the characteristic energies differ from those observed at 4|4E MTBs (Figure~\ref{fig2}b,c). We have additionally observed VB quantization at some tilt GBs, see Figure~S3.

\begin{figure}
	\centering
	\includegraphics[width=162mm]{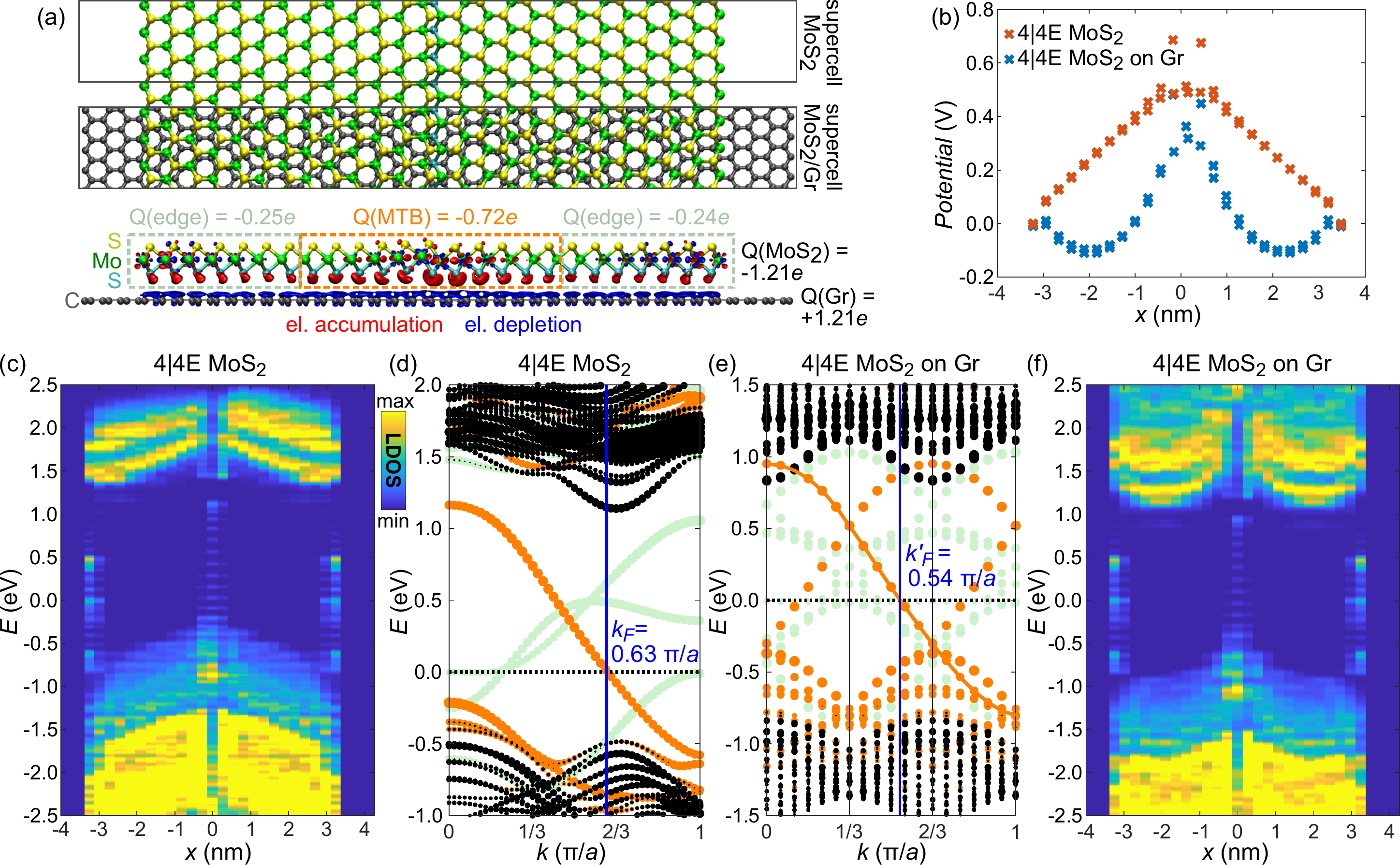}
	\caption{DFT calculations of band bending at 4|4E MTBs in ML-\mo, freestanding or on Gr. (a) DFT geometry of a 4|4E MTB centered in a ribbon of pristine \mo~(top view of upper unit cell) and on Gr (top view of lower unit cell). The side view below illustrates the charge transfer from Gr to the \mo~ribbon by plotting charge difference isosurfaces corresponding to a charge density gain (loss) of $0.002$\,$e$/\AA$^3$ in red (blue). Charge gain or loss is indicated for different sections of the geometry.
	(b) S-atom core potentials perpendicular to the 4|4E MTB in the freestanding \mo~ribbon, red crosses; for the ribbon on Gr, blue crosses. S-atom core potential is set to $0$\,V at the edges in each case.
(c) Projected LDOS for S-atoms perpendicular to the 4|4E MTB in the freestanding \mo~ribbon. LDOS on an arbitrary color scale as indicated.
(d) Band structure of freestanding \mo~ribbon with 4|4E MTB, with different regions highlighted by different colors (MTB: orange; edges: light green; bulk: black).
(e) Band structure of \mo~ribbon on Gr with 4|4E MTB, with different regions highlighted by different colors [as in (d)]. The supercell construction leads to band folding, which is circumvented by using the periodic zone scheme and manually highlighting the MTB band. Fermi-level is set to zero in (d,e), and its crossing of the MTB band marked by a vertical blue line.
(f) Projected LDOS for S-atoms perpendicular to the 4|4E in the \mo~ribbon on Gr. Color scale as in (c).
\label{fig5}}
\end{figure}

\textbf{Density Functional Theory Calculations.} Band bending may be caused by charge accumulation or depletion, but charge density is not directly probeable by STM/STS, typically. To uncover the mechanism of charging, we compare our experimental results to DFT calculations. Figure~\ref{fig5}a shows the geometry of a ML-\mo~ribbon containing a 4|4E-type MTB, the Gr substrate, and the size of the two DFT supercells used to calculate the freestanding \mo~ribbon and the \mo~ribbon on Gr (see Methods). 

We first focus on the freestanding ribbon with a 4|4E MTB. In Figure~\ref{fig5}b, the S-atom core potential variation perpendicular to the 4|4E MTB is plotted (red crosses). The 4|4E MTB is at a higher potential than the ribbon edges, and they are connected through a near-linear slope. In Figure~\ref{fig5}c the corresponding projected LDOS is displayed.
We focus on S atoms because they contribute most to the STS signal, being closest to the STM tip. \textit{Via} STS simulations for pristine MoS$_2$, we verified the S atoms' dominant contribution and note that the LDOS for Mo atoms displays qualitatively the same features.
The variation in the band edge positions in Figure~\ref{fig5}c reflects that of the core potentials. The near-linear potential gradient can be reproduced by describing the \mo~ribbon as a sheet of thickness $d$ and effective dielectric constant $\epsilon$ (compare Ref.~\citenum{ZhangChangjian2014}) containing three parallel wires of line charge $\lambda$, corresponding to the MTB and the two edges. See Figure~S4 for details. Through comparison with our DFT we estimate $\lm = - 0.05$\,$e/a$ at the MTB and $\lambda_{\rm edge} = + 0.025$\,$e/a$ at each edge. Here $e$ is the elementary charge and $a = 3.15$\,\AA~the primitive translation of \mo~along the line. 

The net line charge $\lm = - 0.05$\,$e/a$ is composed of two contributions: (i) the 1D MTB band carries a charge defined by the Fermi wavevector $k_\text{F}$.
The band structure of a freestanding ribbon is shown in Figure~\ref{fig5}d, with MTB and edge bands found inside the bulk gap. Corresponding in-gap states are visible at the MTB and edges in Figure~\ref{fig5}c. For the MTB band $k_\text{F}$ is located at $0.63\frac{\pi}{a}$ and, since the band is hole-like, its filling is $0.37$. Therefore the MTB band carries a band charge $\lmb = -\frac{2e}{\pi}(1-k_\text{F}) =- 0.74$\,$e/a$ (\textit{cf.} $-2\,e/a$ for a full band).
(ii) Evidently, there must be an additional positive charge contribution to the MTB in order to reach the much smaller net line charge of $\lm = - 0.05$\,$e/a$. The ribbon is composed of two \mo~domains of opposite orientation and opposite formal polarization $\mathbf{\pm P}$.\cite{Vanderbilt1993,Gibertini2015}
At the MTB, the abrupt change in polarization $\Delta P$ induces a polarization charge $\lmp$ (discussed in more detail below).
As $\lm = \lmb + \lmp$ we deduce $\lmp = + 0.69$\,$e/a$ on the MTB. The nearly complete compensation of band and polarization charge can be rationalized as follows: the electrostatic energy created by the polarization of the \mo~drives the filling of the 1D MTB and edge bands in order to minimize system energy.\cite{Gibertini2015} In fact, in the limit of infinite ribbon width $\lm$ is exactly zero, as otherwise the energy of the electrostatic field would diverge (`the polar catastrophe').\cite{Gibertini2015} 

When the \mo~ribbon is placed on Gr, a total charge of $-1.21$\,$e$ is depleted from the Gr layer and added to the \mo~ribbon, as determined from Bader charges\cite{Henkelman2006} and illustrated by the charge density difference in the side view of Figure~\ref{fig5}a. Most of this charge is located at the MTB and ribbon edges, with less in the `bulk' regions between. Dividing the \mo~ribbon into MTB and edge regions, we estimate an excess of $-0.72$\,$e$ located at the MTB and $-0.25$\,$e$ at each edge. Considering the $3a$ breadth of the supercell, this equates to linear charge densities of $- 0.24$\,$e/a$ and $- 0.08$\,$e/a$ at MTB and edge, respectively. These transfer charge densities should be added to those of the freestanding \mo~ribbon, \textit{i.e.} a total of $\lm = - 0.29$\,$e$/$a$ is predicted at the 4|4E MTB in \mo~on Gr. Note that, unlike in the \mo~ribbon, the charge distribution in Gr is relatively uniform (compare Figures~\ref{fig5}a and S5) indicating a rather large screening length in Gr. 

The origin of charge transfer from Gr to the \mo~is the work function difference between Gr and ML-\mo. According to DFT and in agreement with experiment, the work function of Gr is lower than the work function of \mo. DFT (experimental) values are around $4.5$\,eV ($4.6$\,eV) for Gr and $5.1$\,eV ($5.2$\,eV) for ML-\mo.\cite{Khomyakov2009,Yu2009,Zhong2016,Choi2014}
Consequently, when \mo~is placed on Gr, the Fermi levels will align through charge transfer. The charge transfer will thus decrease the chemical potential in Gr, and increase the chemical potential in \mo.
If our model for determining $\lambda$ is valid, one should also be able to extract the charge transfer from the change in filling of the MTB band. The band structure from the \mo/Gr supercell is shown in Figure~\ref{fig5}e. Note that due to the incommensurate unit cells of \mo~and Gr the supercell construction must be used, resulting in band folding. This is then displayed in the periodic zone scheme with the MTB band highlighted. We extract the Fermi wavevector of the 1D MTB band $k'_\text{F} = 0.54 \frac {\pi}{a}$ (where $k'_\text{F}$ denotes the presence of Gr). This corresponds to the filling of the hole-like band being increased from $0.37$ to $0.46$, \textit{i.e.} the 1D band charge is increased by $\Delta \lmb = -\frac{2e}{\pi}(k_\text{F}-k'_\text{F}) = - 0.18\,e/a$, in decent agreement with our Bader charge transfer analysis above ($-0.24\,e/a$). In experiment, we measure standing waves along the 4|4E MTB of periodicity $(2.01 \pm 0.04)\,a$ corresponding to a Fermi wavevector $(0.50 \pm 0.01)\frac{\pi}{a}$ (see Ref.~\citenum{Jolie2019}), in agreement with the DFT value $k'_\text{F} = 0.54\frac{\pi}{a}$. Our conclusions here are in line with Kaneko \textit{et al.} \cite{Kaneko2018}, who speculated that the accumulation of charge transferred from the substrate into 1D MTB bands could be the origin of a substantial band bending at MoS$_2$ MTBs.

Figure~\ref{fig5}f shows the projected LDOS of the S atoms perpendicular to the 4|4E MTB for the ribbon on Gr. Large upwards band bending occurs at the MTB and edges in the CBs and VBs, in qualitative agreement with experiment and consistent with the negative line charges on them. Due to the limited cell size the band bending is smaller in magnitude than in experiment; the \mo~band structure does not reach its `bulk' state between the MTB and the edges (compare to the case of an asymmetrically positioned MTB in Figure~S6). The S-atom core potentials for the ribbon with a 4|4E MTB on Gr are plotted in Figure~\ref{fig5}b as blue crosses. They reproduce the variation of the band edge energy with the spatial coordinate. 

\begin{figure}
	\centering
ey    \includegraphics[width=162mm]{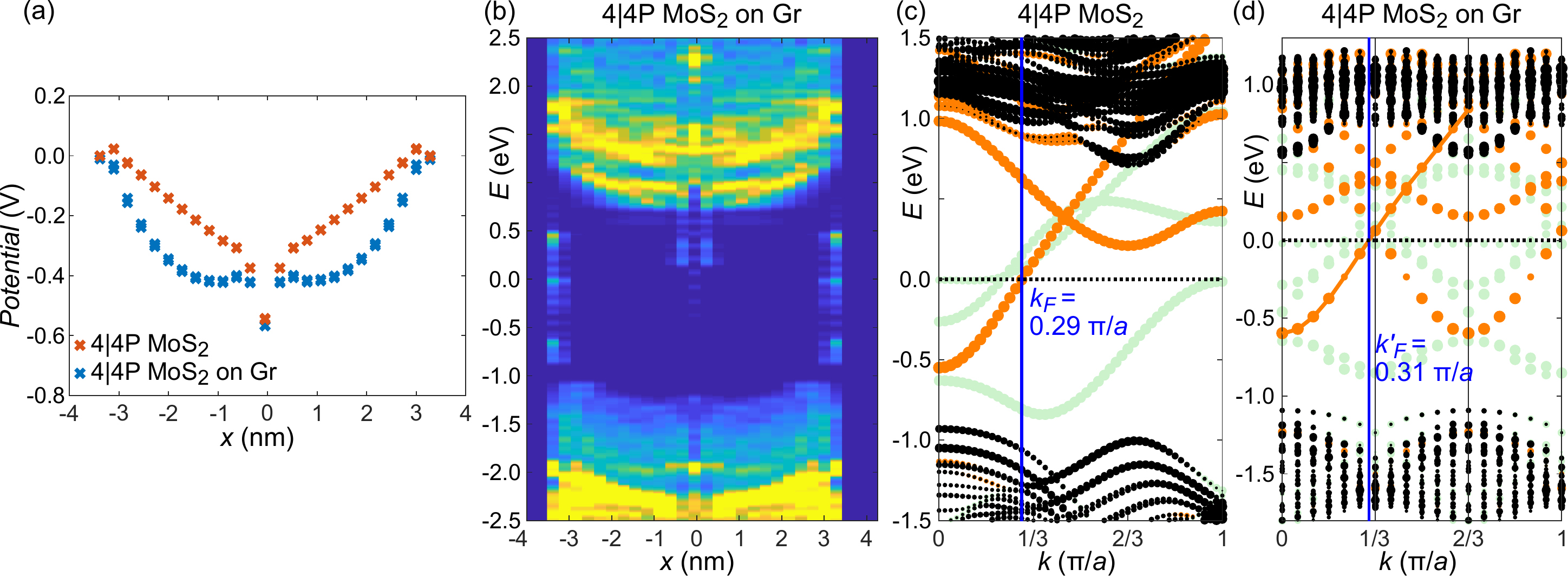}
	\caption{DFT calculations of band bending at 4|4P MTBs in ML-\mo, freestanding or on Gr. Geometry equivalent to that shown in Fig.\,\ref{fig5}a.
(a) S-atom core potentials perpendicular to the 4|4P MTB in the freestanding \mo~ribbon, red crosses; for the ribbon on Gr, blue crosses. S-atom core potential is set to $0$\,V at the edges in each case.
(b) Projected LDOS for S-atoms perpendicular to a 4|4P MTB in a \mo~ribbon on Gr. Color scale as in Fig.\,\ref{fig5}c.
(c) Band structure for freestanding \mo~ribbon with 4|4P MTB.
(d) Band structure for \mo~ribbon on Gr with 4|4P MTB, shown within the periodic zone scheme and manually highlighting the MTB band. 
Fermi-level is set to zero in (c,d); color scheme as in Fig.\,\ref{fig5}d.
          \label{fig6}}
\end{figure}

Corresponding results for the 4|4P-type MTB are shown in Figure~\ref{fig6}.
The S-atom core potentials for the freestanding ribbon, shown in Figure~\ref{fig6}a,
indicate that the MTB is positively charged.
The ribbon on Gr, on the other hand, shows hardly any band bending in the
core potentials or in the projected LDOS (Figure~\ref{fig6}b), as also observed in experiment.
Again, charge transfer from Gr to \mo~takes place when the layers are brought into contact (compare band structures Figure~\ref{fig6}c,d), but results in a nearly neutral net line charge.
This implies that polarization charge and 1D band charge compensate fully ($\lmp \approx -\lmb$), and indeed this conforms with our understanding of the system. 
Assuming the same polarization charge as for the 4|4E MTB would result in an estimate of $\lmb = - 0.69\,e/a$. For the electron-like 1D MTB band ($\lmb = -\frac{2e}{\pi}k_\text{F}$) this corresponds to a Fermi wavevector of $k'_\text{F} = 0.345\frac{\pi}{a}$. This is in excellent agreement with our calculated DFT value $k'_\text{F} = 0.31\frac{\pi}{a}$ (Figure~\ref{fig6}d) and our experimental measurement of $(0.33 \pm 0.01)\frac{\pi}{a}$ [based on $(3.00 \pm 0.04)\,a$ standing wave periodicity in 4|4P MTBs, see Ref.~\citenum{Jolie2019}].

Rephrased, because the charge transfer from Gr fortuitously results in $\lm=0$ for our 4|4P MTB, the measurement of the Fermi wavevector of the 
1D MTB band implies a measurement of the polarization charge $\lmp$. To the authors best knowledge, this is the first direct measurement of a polarization charge. The measured value $\lmp = +0.66\,e/a$ exactly matches $\lmp=+\frac{2}{3}\,e/a$ predicted by Gibertini and Marzari\cite{Gibertini2015} for 4|4P MTBs. More strictly, we are referring here to the `bound charge' which is the sum of $\lambda^{\rm pol}$ and a correction for the changed stochiometry of the interface. The 4|4E MTB was not calculated in Ref.~\citenum{Gibertini2015}, but -- considering the \mo~lattice symmetry and the boundaries' shared Mo$_2$S$_2$ stochiometry -- the 4|4E MTB must possess the same quantized bound and polarization charge as its 4|4P counterpart. 

We note that placing \mo~on Gr as charge reservoir already captures the essence of the processes; the Ir(111) substrate, omitted due to computational limitations, does not change the picture qualitatively. We attribute this to the circumstance that Gr on Ir(111) is almost undoped (Dirac point only 0.1\,eV above $E_\text{F}$); its work function is nearly identical to that of freestanding Gr. However, since Gr offers poor lateral screening, the Ir substrate will affect the lateral screening substantially. 

In our DFT calculations we see no significant lattice strain ($\epsilon<0.5$\%) in the bulk due to MTBs or island edges, ruling out its relevance for band bending. This is consistent with DFT calculations showing that MTBs induce strain only within $\approx0.5$\,nm normal to their axis \cite{Zhao2018,NalinMehta2020}, \textit{i.e.} on a length scale ten times smaller than that of our bending. In previous works TMDC band bending was ascribed to strain at the 1D interfaces \cite{Zhang2018,Ma2017b,Wang2018b,NalinMehta2020}, or a combination of charging and strain \cite{Huang2015,Kobayashi2016,Liu2016,Yan2018}. This may be meaningful for low-symmetry tilt GBs composed of arrays of point dislocations giving rise to large local strains\cite{Huang2015}, or lateral heterostructures with mismatch strain\cite{Zhang2018}, but not for the present case of perfect 4|4E MTBs and straight edges.

\begin{figure}
	\centering
	\includegraphics[width=77mm]{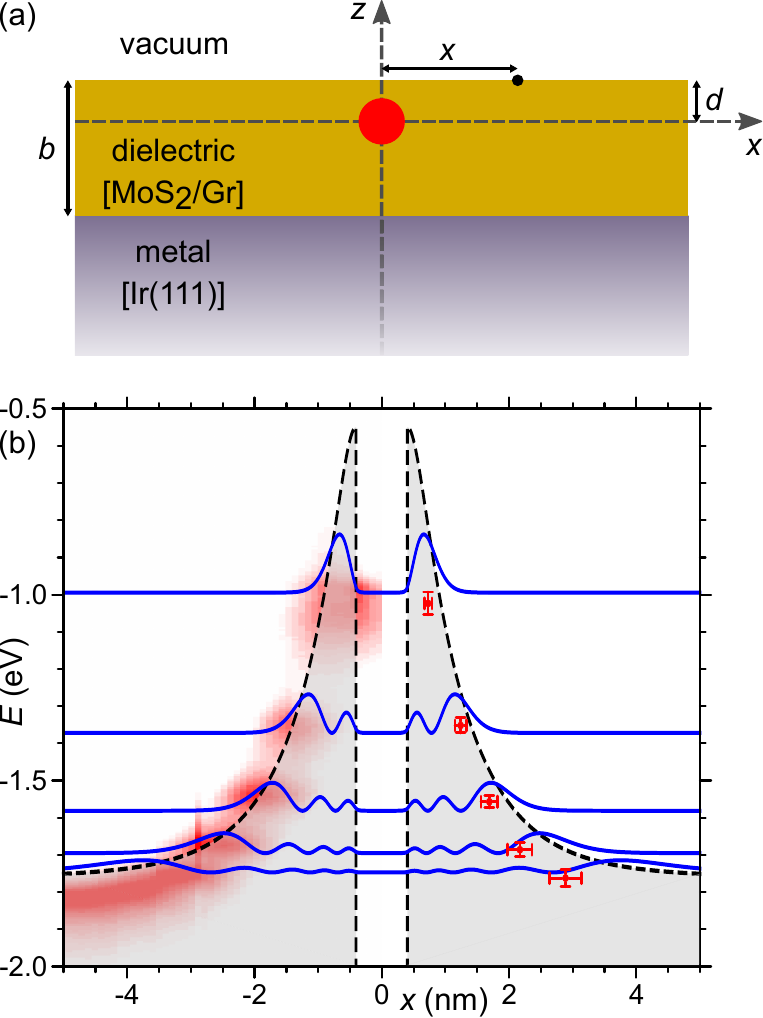}
	\caption{Modeling the VB bending and quantization at a 4|4E MTB (a) Cross-sectional sketch of the model calculating the potential due to a screened, infinite line of charge, see Equation~\ref{eq1}. $b$ is the thickness of the \mo/Gr layer and $d$ the `depth' of the charge below the vacuum interface. $V(x)$ is calculated for a point on the surface, $x$ being the orthogonal distance from the axis of the MTB.
	 (b) Model potential (black) and resulting quantized states (blue) obtained using the values given in the text. The quantized state energy eigenvalues (baselines) are added to their probability densities (displayed as maxima of arbitrary height). For visual comparison, the STS linescan of Figure~\ref{fig2}b (color filtered) and the mean experimental maxima values are shown in red on the left- and right-hand sides, respectively.
		\label{fig7}}
\end{figure}

\textbf{Continuum Modeling.} The potential used in Figure~\ref{fig3}b to reproduce the quantization at the 4|4E MTB was fitted `by hand' and not based on a physical origin. In the following we will use the 4|4E MTB linear charge density $\lm=-0.29\,e/a$ as derived from DFT and physically meaningful system parameters in a simplistic model, in an effort to reproduce the band bending potential and the resultant dominant probability density maxima in $E$ and $x$, as visible in experiment.

The band bending in the \mo~ribbon on Gr/Ir(111) is modeled by a screened electrostatic potential due to an infinite line of charge located at the MTB, sandwiched between vacuum and a perfect metal substrate, as sketched in Figure~\ref{fig7}a. Following the work of Le Quang \textit{et al.} \cite{LeQuang2018}, the MTB charge `generates' image charges in the vacuum and the metal, which themselves generate additional image charges, and so on to the $n$th order. The potential at the \mo~surface at a lateral distance $x$ is then:

\begin{equation}\label{eq1}
V(x)=\frac{4k_\text{e}\lambda}{\epsilon_\text{di}+\epsilon_\text{vac}}\{\ln\frac{d}{\sqrt{x^2+d^2}}+\sum_{n=1}^{\infty}(-1)^n\gamma^{n-1}[\ln\frac{2nb-d}{\sqrt{x^2+(2nb-d)^2}}+\gamma\ln\frac{2nb+d}{\sqrt{x^2+(2nb+d)^2}}]\}+C
\end{equation}

where $k_\text{e}=(4\pi\epsilon_0)^{-1}$, $\gamma=\frac{\epsilon_\text{di}-\epsilon_\text{vac}}{\epsilon_\text{di}+\epsilon_\text{vac}}$, $\epsilon_\text{di}$ and $\epsilon_\text{vac}=1$ are the dielectric constants of the dielectric layer and vacuum respectively, $\lambda$ is the linear charge density at the MTB, $b$ and $d$ are as defined in Figure~\ref{fig7}a, and $C$ is a potential energy offset (as any potential energy is only defined relative to another). The first term in Eq.\ (\ref{eq1}) corresponds to self-screening by the dielectric only, while the second term in Eq.\ (\ref{eq1}), the summation of potentials due to image charges to the $n$th order, is due to metal screening. Finally, at the location of the MTB we add a potential barrier of depth $-5$\,eV and width $0.8$\,nm by hand. We solve the Schr\"{o}dinger equation for positive charge carriers of effective mass $m^*_\text{h}$ in the resultant potential. 

As it is the $\Gamma$-point which is detected in STS in the VB, we use an effective hole mass $m^*_\text{h}=2.7m_\text{e}$ \cite{Peelaers2012,Yu2015}. We consider the charge to be located in the center of the ML-\mo~layer, thus $d=0.31$\,nm. For simplicity, we model the \mo~and Gr (and their van der Waals gaps) together as a single dielectric layer of thickness $b=1.2$\,nm. Gr is only slightly doped in this configuration \cite{Ehlen2018,Pletikosic2009}, and thus expected to screen rather weakly, consistent with the nearly uniform charge distribution in Gr in Figure~\ref{fig5}a. Indeed, considering Gr to be part of the perfect metal substrate leads to an over-estimation of the screening \cite{LeQuang2018}. The dielectric constant of ML-\mo~is anisotropic and non-trivial \cite{ZhangChangjian2014,Laturia2018}, and is complicated further by combination with Gr in our model. It is typically in the range $\epsilon_\text{di}=5-10$, with similar values predicted for a ML-\mo/Gr heterostructure.\cite{Qiu2018}

Setting $\lm = - 0.29\,e/a$, and with only $\epsilon_\text{di}$ and $C$ as variables, we find that values of $\epsilon_\text{di}=6.2$ and $C=-0.55$\,eV yield a very good fit, Figure~\ref{fig7}b. The potential reproduces the lateral positions of the main probability density maxima and  the energy level positions with error less of than $0.03$\,eV, see Table~S1 in Supporting Information.

This continuum modeling neglects the atomistic details of the system, the anisotropy of its screening, and simplifies its layered structure. Nevertheless, it captures the system's essentials. Using literature-based estimates for $b$, $d$, $\epsilon_\text{di}$, $m^*_\text{h}$ and the linear charge density $\lm = - 0.29\,e/a$ obtained by DFT, the modeling reproduces eigenenergies and spatial positions of the quantized VB states at 4|4E MTBs.

\textbf{Discussion.}
Based on literature reports, band bending of the VB and the CB at 1D defects appears to be the rule rather than the exception in TMDC semiconductors \cite{Zhang2014,Huang2015,Wang2018b,Liu2016,Precner2018,Yan2018,Zhang2016,LeQuang2018,Kobayashi2016,Zhang2018}. Our results are in line with this general finding, although the 4|4P MTB does not show band bending. Specifically, at 4|4E MTBs both charge carrier types face a significant potential barrier, with obvious negative implications for the conductivity of the \mo. Electrons face a barrier of at least $0.6$\,eV, while holes become trapped in an even deeper potential well.

A key point emerging from our DFT calculations is that the band bending depends heavily on the substrate, as the substrate's electronic properties -- specifically its work function -- determine the amount of charge transferred. For our case of slightly \textit{p}-doped Gr (Dirac point $0.1$\,eV above $E_\text{F}$\cite{Pletikosic2009}), electron transfer from the substrate to the 1D defects prevails, causing an upward shift of the bands next to 4|4E MTBs and layer edges. By strong \textit{n}- or \textit{p}-doping of Gr, its work function \cite{Schumacher2013NL,Schroder2016}, thus the charge transfer, and consequently the amount of band bending at each 1D line defect is likely to be changed. This could for instance cause upwards or downwards bending also at the 4|4P MTB, for which we consider the absence of band bending in our measurements to be coincidental and related to the specific amount of charge transferred to the system. The sensitive dependence of the band bending on the environment could also partly explain why the measured band bending for MoS$_2$ GBs with similar tilt angles differs vastly \cite{Huang2015,Ly2016,Yan2018,Precner2018}.   

If the nature of the substrate or chemical gating are able to affect band bending, electrostatic gating will do so as well. Thus the magnitude of band bending at a 1D defect is not a fixed quantity, but will change during device operation. On one hand, this will make the effect of 1D defects on charge carrier transport harder to predict. On the other, it could mean that the transport through MTBs and the carrier trapping at MTBs/edges can be controlled or switched \textit{via} gating.

For hole transport through 4|4E MTBs, perhaps an even greater hurdle than bent bands is the repulsive barrier, which is obvious from the complete independence of the quantized VB states on either side of the MTB. Though tunneling transport is likely to take place (in our modeling the barrier width was set to $0.8$\,nm), the transmission will be strongly diminished by the barrier. The barrier is a natural consequence of the broken crystal symmetry at the location of the MTB, causing back-scattering. Though speculation, it is likely that similar barriers for holes exist also for other grain boundaries, \textit{e.g.} at the 4|4P MTB. It thus seems that hole transmission across MTBs is not only suppressed in the energy range of the spin-orbit splitting of the VB due to spin-valley locking, as described by Pulkin and Yazyev \cite{Pulkin2016}, but is globally impaired due to a barrier at the location of the MTB causing backscattering. Indeed Park \textit{et al.}\cite{Park2019} have noted a strong suppression of transmission deep within the VB at 4|4P MTBs. Considering these results, it appears rewarding to investigate hole transport across 4|4E and 4|4P MTBs in more detail, for example with a 4-probe STM. 

In our work, quantization in the VB next to line defects was found whenever a substantial upwards shift of several hundred meV was observed: at 4|4E MTBs, at ML and BL edges, as well as at low symmetry GBs (see Supporting Information). It is evident that such quantization effects should also be present in other, similar systems which display band bending. Consequently the question arises: why has such quantization not been noted in the past? Firstly, we remark that the constant height STS linescan of a BL-\mose~edge by Zhang \textit{et al}. \cite{Zhang2016} (compare Figure 3b of Ref.~\citenum{Zhang2016}) indeed displays step-like features in the upper VB edge, which could be interpreted as signatures of VB quantization in view of the current findings. Next, we note that the VB states were made clearly visible only by constant current STS (or equivalently normalized constant height STS, see Figure~S2), conducted with sufficient resolution and at low temperature.

Based on our DFT calculations we obtain the net line charge $\lm = - 0.29\,e/a$ on the 4|4E MTB in MoS$_2$ on Gr. Due to the marginal doping of Gr on Ir(111), this is also expected to be a reasonable estimate for that on Gr/Ir(111). Using this line charge and realistic system parameters, we were able to fit energy and location of the confined states in the VB accurately with a simple electrostatic model \cite{LeQuang2018}. Though imperfect, the agreement is reasonable and substantiates a net line charge of $\lm \approx-0.29\,e/a$. Our result is similar to those obtained by Le Quang \textit{et al.}\cite{LeQuang2018}, with $\lambda=-0.27\,e/a$ for the edge of a trilayer WSe$_2$ flake, while Kobayashi \textit{et al.}\cite{Kobayashi2016} obtain linear charge densities an order of magnitude larger for a MoS$_2$-WS$_2$ 1D heterojunction interface.

The 2D interfaces of insulating oxide heterostructures can be electrically conductive, in contrast to their bulk constituents. The origin of this conductivity is a matter of debate; a proposed mechanism is the formation of polarization charge at the interface to avoid a so-called `polar-catastrophe'.\cite{Ohtomo2004} The study of analogous polar discontinuities at the 1D interfaces of 2D materials is an emerging field \cite{Gibertini2014,Martinez-Gordillo2015,Park2014}. To the best of our knowledge, the quantitative verification of the polarization charge at the MTBs here represents the first experimental measurement of such interface polarization charge. Considering the complications due to defects and intermixing at 3D hetero-interfaces,\cite{Mannhart2010a,Pai2018} the pristine interfaces of \mo~MTBs may act as a reduced-dimensionality testing ground for polar-charging models.

The well-defined band bending and localization of charges at 4|4E MTBs and edges could find application in optics and optoelectronics. The asymmetry of the bending means a decreasing exciton energy profile towards the defect, possibly leading to an exciton funneling effect as has been achieved through strain-induced bending \cite{Feng2012,Branny2017} -- it might also be possible to control the funneling \textit{via} the aforementioned electrostatic gating. In analogy to TMDC semiconductor point defects having bound excitons \cite{Carozo2017,Dubey2017} and serving as single photon emitters \cite{Zheng2019}, the quantized VB states next to 1D defects in \mo~could also be utilized.

\section{Conclusions}
In summary, we have investigated the electronic landscape perpendicular to line defects in \mo~islands with high-resolution STS. Band bending occurs in a $5$\,nm range at MTBs and ML- as well as BL-island edges, as a result of charge on these 1D defects. Our experiments and DFT calculations develop a fully consistent picture of the defect charging, which is in agreement with literature predictions. \cite{Kaneko2018,Gibertini2015} Namely, the net line charge on MTBs in \mo~is the result of polarization charge, its compensation by 1D MTB band-filling, and additional charge transfer into these in-gap bands due to the substrate's different work function. Our work shows that STS is a viable tool to investigate the charging of domain boundaries. For the 4|4P MTB the absence of band bending implies the net absence of charge on the MTB. Therefore, \textit{via} the STS measurement of the 1D band charge on the MTB, a first direct confirmation of the quantized polarization charge on MTBs was possible, precisely confirming the value of $+\frac{2}{3}\,e/a$ predicted by Gibertini and Marzari\cite{Gibertini2015}. For the 4|4E MTB, the knowledge of the quantized polarization charge together with the measurement of the 1D band charge allows us to estimate the net charge on the MTB, which is in excellent agreement with the prediction of the DFT calculation.

The large ($\approx0.8$\,eV) VB bending normal to the 4|4E-type MTBs on Gr/Ir(111) leads to VB quantization next to it, with a barrier at the MTB creating two independent confining potentials for holes to its left and right. Therefore, hole transport through these MTBs is suppressed. Using a simplistic electrostatic model, with realistic system parameters and the DFT-derived line charge $\lm = - 0.29\,e/a$, the resulting screened potential reproduces band bending accurately. Solving the 1D Schr\"{o}dinger equation for this potential reproduces the eigenenergies and spatial position of the dominant peaks in the probability density. The CB next to 4|4E MTBs displays band bending too, thus CB electron transport is also substantially impaired. Hence this MTB is a considerable barrier to both charge-carrier types. Based on our findings, band bending of the VB and CB normal to the 1D line defects is proposed to be tunable through chemical or electrostatic gating.
 
\section{Methods}

Samples were grown \textit{in situ} in a preparation chamber with a base pressure $<5\times10^{-10}$\,mbar connected to a $5$\,K bath cryostat STM system. 
Ir(111) was cleaned by cycles of $1.5$\,keV Ar$^{+}$ ion erosion and flash annealing to $1550$\,K. Gr is grown on top by room temperature ethylene exposure till saturation and subsequent thermal decomposition at $1370$\,K. The resulting well-oriented Gr islands are grown to a complete single crystal Gr layer through exposure to $2000$\,L ethylene at $1370$\,K~\cite{VanGastel2009}. MoS$_2$ is grown at $300$\,K on Gr/Ir(111) by Mo deposition with a flux of about 
$5 \times 10^{15}$\,atoms\,m$^{-2}$\,s$^{-1}$ in a background elemental sulfur pressure of about $1\times10^{-8}$\,mbar. Subsequently the sample is annealed  for $300$\,s at $1050$\,K in a S pressure of the same magnitude. Compare Ref.\,\citenum{Hall2017}.
 
STM and STS measurements are performed at $T=5$\,K and pressures $<5\times10^{-11}$\,mbar. STS is carried out with the lock-in technique, using a modulation frequency $777$\,Hz and modulation voltage $V_{\text{mod}}=4$\,mV$_{\text{rms}}$. This yields an experimental resolution of $\approx10$\,meV or better \cite{Morgenstern2003}. We use both constant height [recording $(\text{d}I/\text{d}V)_Z$] and constant current [$(\text{d}I/\text{d}V)_I$] STS, where $I$ is the tunneling current, $V$ the bias voltage, and $Z$ the height of the tip above the sample. In both modes $\text{d}I/\text{d}V$ is recorded while $V$ is ramped. For further explanation see Ref.~\citenum{Murray2019}. A linescan consists of a sequence of spectra taken at most $1$\,\AA~apart along a straight line. In each spectrum, data points are taken every $5-20$\,meV depending on the desired resolution. The path of the linescan may not be perfectly normal to the 1D defect, but the resulting error in $x$ can be neglected \textemdash~in the scan of Figure~\ref{fig1}c, for example, this amounts to a factor cos$(1.9^\circ)=0.999$.  The color plots of the linescans show raw data; no interpolation or smoothing is involved.  

All density functional theory calculations were carried out using Vienna \textit{Ab Initio} Simulation Package (VASP) \cite{Kresse1993,Kresse1996}. The plane wave cutoff was set to 400 eV throughout.
The atomic structure for the ML-MoS$_2$ ribbon with MTB and unpassivated Mo-edges on Gr, as shown in Fig.\ \ref{fig5}, was constructed as follows.
While the strain in the direction perpendicular to the MTB ($\mathbf{a}_1$) is naturally released in the ribbon geometry, a suitable supercell needs to be constructed to release strain in the direction parallel to the MTB ($\mathbf{a}_2$).
Since $3a$ of MoS$_2$ fairly closely matches with $4a$ of Gr, we use a MoS$_2$ layer consisting of 3$\times$(12+12) unit cells (12 units in each side of the MTB) and a Gr layer consisting of 4$\times$40 unit cells.
The supercell is hexagonal and thus the distance between the MTB and the edge of the ribbon is $6\sqrt{3}a \approx 33.3$ {\AA}. The lattice constant of MoS$_2$ along $\bf{a}_2$ is fixed to that of pristine ML-MoS$_2$, which yields a Gr lattice constant $2.385$\,{\AA}, not very far from the optimized value of $2.468$\,{\AA}. The lattice constant of Gr along $\bf{a}_1$ is fixed to that of pristine Gr. The carbon atom z-coordinates are fixed to prevent buckling.
We use the exchange-correlation functional of Perdew, Burke, and Ernzerhof (PBE) \cite{Perdew1996}, augmented with Grimme's corrections (-D2) for the van der Waals interactions \cite{Grimme2006}. A 4$\times$1$\times$1 k-point mesh is used during ionic relaxation. The density of states is evaluated with a 12$\times$1 mesh, \textit{i.e.} corresponding to a 36$\times$1 mesh for MoS$_2$ ribbon and 48$\times$40 for Gr.

\begin{acknowledgement}
This  work  was  funded  by  the  Deutsche  Forschungsgemeinschaft (DFG, German Research Foundation), CRC 1238 (project number 277146847, subprojects A01, B06 and C02). Support from the German Academic Exchange Service DAAD \textit{via} PPP Finland 'MODEST', project ID 57458732 is gratefully acknowledged. H.P.K. acknowledges financial support from the Academy of Finland through Project No. 311058 and CSC-IT Center for Science Ltd. for generous grants of computer time. A.V.K acknowledges funding from DFG through project KR 48661-2 (406129719). The authors thank Carsten Busse and Mahdi Ghorbani-Asl for useful discussions.

\end{acknowledgement}

\begin{suppinfo}
	
	The Supporting Information is available free of charge at: 
	
	Supplementary STM/STS measurements of 4|4P and 4|4E MTBs, an explanation of the analysis of quantized state locations, an example of VB quantization occurring at a tilt-angle GB, charged wire modeling used to estimate the net linear charge of 1D defects in a freestanding \mo~layer, additional DFT calculations of the 4|4E MTB, and fit values resulting from the electrostatic continuum model.
	
\end{suppinfo}	

\bibliography{lit}

\providecommand{\latin}[1]{#1}
\makeatletter
\providecommand{\doi}
  {\begingroup\let\do\@makeother\dospecials
  \catcode`\{=1 \catcode`\}=2 \doi@aux}
\providecommand{\doi@aux}[1]{\endgroup\texttt{#1}}
\makeatother
\providecommand*\mcitethebibliography{\thebibliography}
\csname @ifundefined\endcsname{endmcitethebibliography}
  {\let\endmcitethebibliography\endthebibliography}{}
\begin{mcitethebibliography}{68}
\providecommand*\natexlab[1]{#1}
\providecommand*\mciteSetBstSublistMode[1]{}
\providecommand*\mciteSetBstMaxWidthForm[2]{}
\providecommand*\mciteBstWouldAddEndPuncttrue
  {\def\EndOfBibitem{\unskip.}}
\providecommand*\mciteBstWouldAddEndPunctfalse
  {\let\EndOfBibitem\relax}
\providecommand*\mciteSetBstMidEndSepPunct[3]{}
\providecommand*\mciteSetBstSublistLabelBeginEnd[3]{}
\providecommand*\EndOfBibitem{}
\mciteSetBstSublistMode{f}
\mciteSetBstMaxWidthForm{subitem}{(\alph{mcitesubitemcount})}
\mciteSetBstSublistLabelBeginEnd
  {\mcitemaxwidthsubitemform\space}
  {\relax}
  {\relax}

\bibitem[{Van Der Zande} \latin{et~al.}(2013){Van Der Zande}, Huang, Chenet,
  Berkelbach, You, Lee, Heinz, Reichman, Muller, and Hone]{VanDerZande2013}
{Van Der Zande},~A.~M.; Huang,~P.~Y.; Chenet,~D.~A.; Berkelbach,~T.~C.;
  You,~Y.; Lee,~G.~H.; Heinz,~T.~F.; Reichman,~D.~R.; Muller,~D.~A.;
  Hone,~J.~C. {Grains and Grain Boundaries in Highly Crystalline Monolayer
  Molybdenum Disulphide}. \emph{Nat. Mater.} \textbf{2013}, \emph{12},
  554--561\relax
\mciteBstWouldAddEndPuncttrue
\mciteSetBstMidEndSepPunct{\mcitedefaultmidpunct}
{\mcitedefaultendpunct}{\mcitedefaultseppunct}\relax
\EndOfBibitem
\bibitem[Najmaei \latin{et~al.}(2014)Najmaei, Amani, Chin, Liu, Birdwell,
  O'Regan, Ajayan, Dubey, and Lou]{Najmaei2014}
Najmaei,~S.; Amani,~M.; Chin,~M.~L.; Liu,~Z.; Birdwell,~A.~G.; O'Regan,~T.~P.;
  Ajayan,~P.~M.; Dubey,~M.; Lou,~J. {Electrical Transport Properties of
  Polycrystalline Monolayer Molybdenum Disulfide}. \emph{ACS Nano}
  \textbf{2014}, \emph{8}, 7930--7937\relax
\mciteBstWouldAddEndPuncttrue
\mciteSetBstMidEndSepPunct{\mcitedefaultmidpunct}
{\mcitedefaultendpunct}{\mcitedefaultseppunct}\relax
\EndOfBibitem
\bibitem[Du \latin{et~al.}(2016)Du, Yu, Xie, Wu, Wang, Lu, Liao, Meng, Zhao,
  Zhang, Zhu, Chen, Wang, Yang, Shi, and Zhang]{Du2016}
Du,~L.; Yu,~H.; Xie,~L.; Wu,~S.; Wang,~S.; Lu,~X.; Liao,~M.; Meng,~J.;
  Zhao,~J.; Zhang,~J.; Zhu,~J.; Chen,~P.; Wang,~G.; Yang,~R.; Shi,~D.;
  Zhang,~G. {The Effect of Twin Grain Boundary Tuned by Temperature on the
  Electrical Transport Properties of Monolayer MoS$_2$}. \emph{Crystals}
  \textbf{2016}, \emph{6}, 115\relax
\mciteBstWouldAddEndPuncttrue
\mciteSetBstMidEndSepPunct{\mcitedefaultmidpunct}
{\mcitedefaultendpunct}{\mcitedefaultseppunct}\relax
\EndOfBibitem
\bibitem[Ly \latin{et~al.}(2016)Ly, Perello, Zhao, Deng, Kim, Han, Chae, Jeong,
  and Lee]{Ly2016}
Ly,~T.~H.; Perello,~D.~J.; Zhao,~J.; Deng,~Q.; Kim,~H.; Han,~G.~H.;
  Chae,~S.~H.; Jeong,~H.~Y.; Lee,~Y.~H. {Misorientation-Angle-Dependent
  Electrical Transport across Molybdenum Disulfide Grain Boundaries}.
  \emph{Nat. Commun.} \textbf{2016}, \emph{7}, 10426\relax
\mciteBstWouldAddEndPuncttrue
\mciteSetBstMidEndSepPunct{\mcitedefaultmidpunct}
{\mcitedefaultendpunct}{\mcitedefaultseppunct}\relax
\EndOfBibitem
\bibitem[Kimoto(2015)]{Kimoto2015}
Kimoto,~T. {Material Science and Device Physics in SiC Technology for
  High-Voltage Power Devices}. \emph{Jpn. J. Appl. Phys.} \textbf{2015},
  \emph{54}, 040103\relax
\mciteBstWouldAddEndPuncttrue
\mciteSetBstMidEndSepPunct{\mcitedefaultmidpunct}
{\mcitedefaultendpunct}{\mcitedefaultseppunct}\relax
\EndOfBibitem
\bibitem[Skowronski and Ha(2006)Skowronski, and Ha]{Skowronski2006}
Skowronski,~M.; Ha,~S. {Degradation of Hexagonal Silicon-Carbide-Based Bipolar
  Devices}. \emph{J. Appl. Phys.} \textbf{2006}, \emph{99}, 011101\relax
\mciteBstWouldAddEndPuncttrue
\mciteSetBstMidEndSepPunct{\mcitedefaultmidpunct}
{\mcitedefaultendpunct}{\mcitedefaultseppunct}\relax
\EndOfBibitem
\bibitem[Huang \latin{et~al.}(2015)Huang, Chen, Zhang, Quek, Chen, Li, Hsu,
  Chang, Zheng, Chen, and Wee]{Huang2015}
Huang,~Y.~L.; Chen,~Y.; Zhang,~W.; Quek,~S.~Y.; Chen,~C.-H.; Li,~L.-J.;
  Hsu,~W.-T.; Chang,~W.-H.; Zheng,~Y.~J.; Chen,~W.; Wee,~A. T.~S. {Bandgap
  Tunability at Single-Layer Molybdenum Disulphide Grain Boundaries}.
  \emph{Nat. Commun.} \textbf{2015}, \emph{6}, 6298\relax
\mciteBstWouldAddEndPuncttrue
\mciteSetBstMidEndSepPunct{\mcitedefaultmidpunct}
{\mcitedefaultendpunct}{\mcitedefaultseppunct}\relax
\EndOfBibitem
\bibitem[Wang \latin{et~al.}(2018)Wang, Yu, Tao, Xiao, Fan, Zhang, Liao, Guo,
  Shi, Du, Zhang, and Gao]{Wang2018b}
Wang,~D.; Yu,~H.; Tao,~L.; Xiao,~W.; Fan,~P.; Zhang,~T.; Liao,~M.; Guo,~W.;
  Shi,~D.; Du,~S.; Zhang,~G.; Gao,~H. {Bandgap Broadening at Grain Boundaries
  in Single-Layer MoS$_2$}. \emph{Nano Res.} \textbf{2018}, \emph{11},
  6102--6109\relax
\mciteBstWouldAddEndPuncttrue
\mciteSetBstMidEndSepPunct{\mcitedefaultmidpunct}
{\mcitedefaultendpunct}{\mcitedefaultseppunct}\relax
\EndOfBibitem
\bibitem[Yan \latin{et~al.}(2018)Yan, Dong, Li, and Li]{Yan2018}
Yan,~C.; Dong,~X.; Li,~C.~H.; Li,~L. {Charging Effect at Grain Boundaries of
  MoS$_2$}. \emph{Nanotechnology} \textbf{2018}, \emph{29}, 195704\relax
\mciteBstWouldAddEndPuncttrue
\mciteSetBstMidEndSepPunct{\mcitedefaultmidpunct}
{\mcitedefaultendpunct}{\mcitedefaultseppunct}\relax
\EndOfBibitem
\bibitem[Liu \latin{et~al.}(2016)Liu, Balla, Bergeron, and Hersam]{Liu2016}
Liu,~X.; Balla,~I.; Bergeron,~H.; Hersam,~M.~C. {Point Defects and Grain
  Boundaries in Rotationally Commensurate MoS$_2$ on Epitaxial Graphene}.
  \emph{J. Phys. Chem. C} \textbf{2016}, \emph{120}, 20798--20805\relax
\mciteBstWouldAddEndPuncttrue
\mciteSetBstMidEndSepPunct{\mcitedefaultmidpunct}
{\mcitedefaultendpunct}{\mcitedefaultseppunct}\relax
\EndOfBibitem
\bibitem[Precner \latin{et~al.}(2018)Precner, Polakovi{\'{c}}, Qiao, Trainer,
  Putilov, {Di Giorgio}, Cone, Zhu, Xi, Iavarone, and Karapetrov]{Precner2018}
Precner,~M.; Polakovi{\'{c}},~T.; Qiao,~Q.; Trainer,~D.~J.; Putilov,~A.~V.; {Di
  Giorgio},~C.; Cone,~I.; Zhu,~Y.; Xi,~X.~X.; Iavarone,~M.; Karapetrov,~G.
  {Evolution of Metastable Defects and Its Effect on the Electronic Properties
  of MoS$_2$ Films}. \emph{Sci. Rep.} \textbf{2018}, \emph{8}, 6724\relax
\mciteBstWouldAddEndPuncttrue
\mciteSetBstMidEndSepPunct{\mcitedefaultmidpunct}
{\mcitedefaultendpunct}{\mcitedefaultseppunct}\relax
\EndOfBibitem
\bibitem[Zhang \latin{et~al.}(2014)Zhang, Johnson, Hsu, Li, and
  Shih]{Zhang2014}
Zhang,~C.; Johnson,~A.; Hsu,~C.-L.; Li,~L.-J.; Shih,~C.-K. {Direct Imaging of
  Band Profile in Single Layer MoS$_2$ on Graphite: Quasiparticle Energy Gap,
  Metallic Edge States, and Edge Band Bending}. \emph{Nano Lett.}
  \textbf{2014}, \emph{14}, 2443--2447\relax
\mciteBstWouldAddEndPuncttrue
\mciteSetBstMidEndSepPunct{\mcitedefaultmidpunct}
{\mcitedefaultendpunct}{\mcitedefaultseppunct}\relax
\EndOfBibitem
\bibitem[Zhang \latin{et~al.}(2016)Zhang, Chen, Huang, Wu, Li, Yao, Tersoff,
  and Shih]{Zhang2016}
Zhang,~C.; Chen,~Y.; Huang,~J.-K.; Wu,~X.; Li,~L.-J.; Yao,~W.; Tersoff,~J.;
  Shih,~C.-K. {Visualizing Band Offsets and Edge States in Bilayer-Monolayer
  Transition Metal Dichalcogenides Lateral Heterojunction}. \emph{Nat. Commun.}
  \textbf{2016}, \emph{7}, 10349\relax
\mciteBstWouldAddEndPuncttrue
\mciteSetBstMidEndSepPunct{\mcitedefaultmidpunct}
{\mcitedefaultendpunct}{\mcitedefaultseppunct}\relax
\EndOfBibitem
\bibitem[{Le Quang} \latin{et~al.}(2018){Le Quang}, Nogajewski, Potemski, Dau,
  Jamet, Mallet, and Veuillen]{LeQuang2018}
{Le Quang},~T.; Nogajewski,~K.; Potemski,~M.; Dau,~M.~T.; Jamet,~M.;
  Mallet,~P.; Veuillen,~J.-Y. {Band-Bending Induced by Charged Defects and
  Edges of Atomically Thin Transition Metal Dichalcogenide Films}. \emph{2D
  Mater.} \textbf{2018}, \emph{5}, 035034\relax
\mciteBstWouldAddEndPuncttrue
\mciteSetBstMidEndSepPunct{\mcitedefaultmidpunct}
{\mcitedefaultendpunct}{\mcitedefaultseppunct}\relax
\EndOfBibitem
\bibitem[Kobayashi \latin{et~al.}(2016)Kobayashi, Yoshida, Sakurada, Takashima,
  Yamamoto, Saito, Konabe, Taniguchi, Watanabe, Maniwa, Takeuchi, Shigekawa,
  and Miyata]{Kobayashi2016}
Kobayashi,~Y.; Yoshida,~S.; Sakurada,~R.; Takashima,~K.; Yamamoto,~T.;
  Saito,~T.; Konabe,~S.; Taniguchi,~T.; Watanabe,~K.; Maniwa,~Y.; Takeuchi,~O.;
  Shigekawa,~H.; Miyata,~Y. {Modulation of Electrical Potential and
  Conductivity in an Atomic-Layer Semiconductor Heterojunction}. \emph{Sci.
  Rep.} \textbf{2016}, \emph{6}, 31223\relax
\mciteBstWouldAddEndPuncttrue
\mciteSetBstMidEndSepPunct{\mcitedefaultmidpunct}
{\mcitedefaultendpunct}{\mcitedefaultseppunct}\relax
\EndOfBibitem
\bibitem[Zhang \latin{et~al.}(2018)Zhang, Li, Tersoff, Han, Su, Li, Muller, and
  Shih]{Zhang2018}
Zhang,~C.; Li,~M.-Y.; Tersoff,~J.; Han,~Y.; Su,~Y.; Li,~L.-J.; Muller,~D.~A.;
  Shih,~C.-K. {Strain Distributions and Their Influence on Electronic
  Structures of WSe$_2$-MoS$_2$ Laterally Strained Heterojunctions}. \emph{Nat.
  Nanotechnol.} \textbf{2018}, \emph{13}, 152--158\relax
\mciteBstWouldAddEndPuncttrue
\mciteSetBstMidEndSepPunct{\mcitedefaultmidpunct}
{\mcitedefaultendpunct}{\mcitedefaultseppunct}\relax
\EndOfBibitem
\bibitem[Ma \latin{et~al.}(2017)Ma, Kolekar, {Coy Diaz}, Aprojanz, Miccoli,
  Tegenkamp, and Batzill]{Ma2017b}
Ma,~Y.; Kolekar,~S.; {Coy Diaz},~H.; Aprojanz,~J.; Miccoli,~I.; Tegenkamp,~C.;
  Batzill,~M. {Metallic Twin Grain Boundaries Embedded in MoSe$_2$ Monolayers
  Grown by Molecular Beam Epitaxy}. \emph{ACS Nano} \textbf{2017}, \emph{11},
  5130--5139\relax
\mciteBstWouldAddEndPuncttrue
\mciteSetBstMidEndSepPunct{\mcitedefaultmidpunct}
{\mcitedefaultendpunct}{\mcitedefaultseppunct}\relax
\EndOfBibitem
\bibitem[Kaneko and Saito(2018)Kaneko, and Saito]{Kaneko2018}
Kaneko,~T.; Saito,~R. {Origin of Band Bending at Domain Boundaries of MoS$_2$:
  First-Principles Study}. \emph{Jpn. J. Appl. Phys.} \textbf{2018}, \emph{57},
  04FP09\relax
\mciteBstWouldAddEndPuncttrue
\mciteSetBstMidEndSepPunct{\mcitedefaultmidpunct}
{\mcitedefaultendpunct}{\mcitedefaultseppunct}\relax
\EndOfBibitem
\bibitem[Hall \latin{et~al.}(2018)Hall, Pieli{\'{c}}, Murray, Jolie, Wekking,
  Busse, Kralj, and Michely]{Hall2017}
Hall,~J.; Pieli{\'{c}},~B.; Murray,~C.; Jolie,~W.; Wekking,~T.; Busse,~C.;
  Kralj,~M.; Michely,~T. {Molecular Beam Epitaxy of Quasi-Freestanding
  Transition Metal Disulphide Monolayers on van der Waals Substrates: A Growth
  Study}. \emph{2D Mater.} \textbf{2018}, \emph{5}, 025005\relax
\mciteBstWouldAddEndPuncttrue
\mciteSetBstMidEndSepPunct{\mcitedefaultmidpunct}
{\mcitedefaultendpunct}{\mcitedefaultseppunct}\relax
\EndOfBibitem
\bibitem[Coraux \latin{et~al.}(2008)Coraux, N'Diaye, Busse, and
  Michely]{Coraux2008}
Coraux,~J.; N'Diaye,~A.~T.; Busse,~C.; Michely,~T. {Structural Coherency of
  Graphene on Ir(111)}. \emph{Nano Lett.} \textbf{2008}, \emph{8},
  565--570\relax
\mciteBstWouldAddEndPuncttrue
\mciteSetBstMidEndSepPunct{\mcitedefaultmidpunct}
{\mcitedefaultendpunct}{\mcitedefaultseppunct}\relax
\EndOfBibitem
\bibitem[Komsa and Krasheninnikov(2017)Komsa, and Krasheninnikov]{Komsa2017}
Komsa,~H.-P.; Krasheninnikov,~A.~V. {Engineering the Electronic Properties of
  Two-Dimensional Transition Metal Dichalcogenides by Introducing Mirror Twin
  Boundaries}. \emph{Adv. Electron. Mater.} \textbf{2017}, \emph{3},
  1600468\relax
\mciteBstWouldAddEndPuncttrue
\mciteSetBstMidEndSepPunct{\mcitedefaultmidpunct}
{\mcitedefaultendpunct}{\mcitedefaultseppunct}\relax
\EndOfBibitem
\bibitem[Jolie \latin{et~al.}(2019)Jolie, Murray, Wei{\ss}, Hall, Portner,
  Atodiresei, Krasheninnikov, Busse, Komsa, Rosch, and Michely]{Jolie2019}
Jolie,~W.; Murray,~C.; Wei{\ss},~P.~S.; Hall,~J.; Portner,~F.; Atodiresei,~N.;
  Krasheninnikov,~A.~V.; Busse,~C.; Komsa,~H.-P.; Rosch,~A.; Michely,~T.
  {Tomonaga-Luttinger Liquid in a Box: Electrons Confined within MoS$_2$
  Mirror-Twin Boundaries}. \emph{Phys. Rev. X} \textbf{2019}, \emph{9},
  011055\relax
\mciteBstWouldAddEndPuncttrue
\mciteSetBstMidEndSepPunct{\mcitedefaultmidpunct}
{\mcitedefaultendpunct}{\mcitedefaultseppunct}\relax
\EndOfBibitem
\bibitem[Zhou \latin{et~al.}(2013)Zhou, Zou, Najmaei, Liu, Shi, Kong, Lou,
  Ajayan, Yakobson, and Idrobo]{Zhou2013}
Zhou,~W.; Zou,~X.; Najmaei,~S.; Liu,~Z.; Shi,~Y.; Kong,~J.; Lou,~J.;
  Ajayan,~P.~M.; Yakobson,~B.~I.; Idrobo,~J.-C. {Intrinsic Structural Defects
  in Monolayer Molybdenum Disulfide}. \emph{Nano Lett.} \textbf{2013},
  \emph{13}, 2615--2622\relax
\mciteBstWouldAddEndPuncttrue
\mciteSetBstMidEndSepPunct{\mcitedefaultmidpunct}
{\mcitedefaultendpunct}{\mcitedefaultseppunct}\relax
\EndOfBibitem
\bibitem[Zou \latin{et~al.}(2013)Zou, Liu, and Yakobson]{Zou2013}
Zou,~X.; Liu,~Y.; Yakobson,~B.~I. {Predicting Dislocations and Grain Boundaries
  in Two-Dimensional Metal-Disulfides from the First Principles}. \emph{Nano
  Lett.} \textbf{2013}, \emph{13}, 253--258\relax
\mciteBstWouldAddEndPuncttrue
\mciteSetBstMidEndSepPunct{\mcitedefaultmidpunct}
{\mcitedefaultendpunct}{\mcitedefaultseppunct}\relax
\EndOfBibitem
\bibitem[Zou and Yakobson(2015)Zou, and Yakobson]{Zou2015}
Zou,~X.; Yakobson,~B.~I. {Metallic High-Angle Grain Boundaries in Monolayer
  Polycrystalline WS$_2$}. \emph{Small} \textbf{2015}, \emph{11},
  4503--4507\relax
\mciteBstWouldAddEndPuncttrue
\mciteSetBstMidEndSepPunct{\mcitedefaultmidpunct}
{\mcitedefaultendpunct}{\mcitedefaultseppunct}\relax
\EndOfBibitem
\bibitem[Batzill(2018)]{Batzill2018}
Batzill,~M. {Mirror Twin Grain Boundaries in Molybdenum Dichalcogenides}.
  \emph{J. Phys. Condens. Matter} \textbf{2018}, \emph{30}, 493001\relax
\mciteBstWouldAddEndPuncttrue
\mciteSetBstMidEndSepPunct{\mcitedefaultmidpunct}
{\mcitedefaultendpunct}{\mcitedefaultseppunct}\relax
\EndOfBibitem
\bibitem[Murray \latin{et~al.}(2019)Murray, Jolie, Fischer, Hall, van Efferen,
  Ehlen, Gr{\"{u}}neis, Busse, and Michely]{Murray2019}
Murray,~C.; Jolie,~W.; Fischer,~J.~A.; Hall,~J.; van Efferen,~C.; Ehlen,~N.;
  Gr{\"{u}}neis,~A.; Busse,~C.; Michely,~T. {Comprehensive Tunneling
  Spectroscopy of Quasifreestanding MoS$_2$ on Graphene on Ir(111)}.
  \emph{Phys. Rev. B} \textbf{2019}, \emph{99}, 115434\relax
\mciteBstWouldAddEndPuncttrue
\mciteSetBstMidEndSepPunct{\mcitedefaultmidpunct}
{\mcitedefaultendpunct}{\mcitedefaultseppunct}\relax
\EndOfBibitem
\bibitem[Yu \latin{et~al.}(2016)Yu, Kutana, and Yakobson]{Yu2016}
Yu,~H.; Kutana,~A.; Yakobson,~B.~I. {Carrier Delocalization in Two-Dimensional
  Coplanar \textit{p}-\textit{n} Junctions of Graphene and Metal
  Dichalcogenides}. \emph{Nano Lett.} \textbf{2016}, \emph{16},
  5032--5036\relax
\mciteBstWouldAddEndPuncttrue
\mciteSetBstMidEndSepPunct{\mcitedefaultmidpunct}
{\mcitedefaultendpunct}{\mcitedefaultseppunct}\relax
\EndOfBibitem
\bibitem[Zhang \latin{et~al.}(2015)Zhang, Chen, Johnson, Li, Li, Mende,
  Feenstra, and Shih]{Zhang2015}
Zhang,~C.; Chen,~Y.; Johnson,~A.; Li,~M.-Y.; Li,~L.-J.; Mende,~P.~C.;
  Feenstra,~R.~M.; Shih,~C.-K. {Probing Critical Point Energies of Transition
  Metal Dichalcogenides: Surprising Indirect Gap of Single Layer WSe$_2$}.
  \emph{Nano Lett.} \textbf{2015}, \emph{15}, 6494--6500\relax
\mciteBstWouldAddEndPuncttrue
\mciteSetBstMidEndSepPunct{\mcitedefaultmidpunct}
{\mcitedefaultendpunct}{\mcitedefaultseppunct}\relax
\EndOfBibitem
\bibitem[Stroscio and Kaiser(1993)Stroscio, and Kaiser]{Stroscio1993a}
Stroscio,~J.~A.; Kaiser,~W.~J. \emph{{Scanning Tunneling Microscopy}}, 27th
  ed.; Academic Press: San Diego, 1993; pp 102--147\relax
\mciteBstWouldAddEndPuncttrue
\mciteSetBstMidEndSepPunct{\mcitedefaultmidpunct}
{\mcitedefaultendpunct}{\mcitedefaultseppunct}\relax
\EndOfBibitem
\bibitem[Zhang \latin{et~al.}(2014)Zhang, Wang, Chan, Manolatou, and
  Rana]{ZhangChangjian2014}
Zhang,~C.; Wang,~H.; Chan,~W.; Manolatou,~C.; Rana,~F. {Absorption of Light by
  Excitons and Trions in Monolayers of Metal Dichalcogenide MoS$_2$:
  Experiments and Theory}. \emph{Phys. Rev. B} \textbf{2014}, \emph{89},
  205436\relax
\mciteBstWouldAddEndPuncttrue
\mciteSetBstMidEndSepPunct{\mcitedefaultmidpunct}
{\mcitedefaultendpunct}{\mcitedefaultseppunct}\relax
\EndOfBibitem
\bibitem[Vanderbilt and King-Smith(1993)Vanderbilt, and
  King-Smith]{Vanderbilt1993}
Vanderbilt,~D.; King-Smith,~R.~D. {Electric Polarization as a Bulk Quantity and
  Its Relation to Surface Charge}. \emph{Phys. Rev. B} \textbf{1993},
  \emph{48}, 4442--4455\relax
\mciteBstWouldAddEndPuncttrue
\mciteSetBstMidEndSepPunct{\mcitedefaultmidpunct}
{\mcitedefaultendpunct}{\mcitedefaultseppunct}\relax
\EndOfBibitem
\bibitem[Gibertini and Marzari(2015)Gibertini, and Marzari]{Gibertini2015}
Gibertini,~M.; Marzari,~N. {Emergence of One-Dimensional Wires of Free Carriers
  in Transition-Metal-Dichalcogenide Nanostructures}. \emph{Nano Lett.}
  \textbf{2015}, \emph{15}, 6229--6238\relax
\mciteBstWouldAddEndPuncttrue
\mciteSetBstMidEndSepPunct{\mcitedefaultmidpunct}
{\mcitedefaultendpunct}{\mcitedefaultseppunct}\relax
\EndOfBibitem
\bibitem[Henkelman \latin{et~al.}(2006)Henkelman, Arnaldsson, and
  J{\'{o}}nsson]{Henkelman2006}
Henkelman,~G.; Arnaldsson,~A.; J{\'{o}}nsson,~H. {A Fast and Robust Algorithm
  for Bader Decomposition of Charge Density}. \emph{Comput. Mater. Sci.}
  \textbf{2006}, \emph{36}, 354--360\relax
\mciteBstWouldAddEndPuncttrue
\mciteSetBstMidEndSepPunct{\mcitedefaultmidpunct}
{\mcitedefaultendpunct}{\mcitedefaultseppunct}\relax
\EndOfBibitem
\bibitem[Khomyakov \latin{et~al.}(2009)Khomyakov, Giovannetti, Rusu, Brocks,
  van~den Brink, and Kelly]{Khomyakov2009}
Khomyakov,~P.~A.; Giovannetti,~G.; Rusu,~P.~C.; Brocks,~G.; van~den Brink,~J.;
  Kelly,~P.~J. {First-Principles Study of the Interaction and Charge Transfer
  between Graphene and Metals}. \emph{Phys. Rev. B} \textbf{2009}, \emph{79},
  195425\relax
\mciteBstWouldAddEndPuncttrue
\mciteSetBstMidEndSepPunct{\mcitedefaultmidpunct}
{\mcitedefaultendpunct}{\mcitedefaultseppunct}\relax
\EndOfBibitem
\bibitem[Yu \latin{et~al.}(2009)Yu, Zhao, Ryu, Brus, Kim, and Kim]{Yu2009}
Yu,~Y.-J.; Zhao,~Y.; Ryu,~S.; Brus,~L.~E.; Kim,~K.~S.; Kim,~P. {Tuning the
  Graphene Work Function by Electric Field Effect}. \emph{Nano Lett.}
  \textbf{2009}, \emph{9}, 3430--3434\relax
\mciteBstWouldAddEndPuncttrue
\mciteSetBstMidEndSepPunct{\mcitedefaultmidpunct}
{\mcitedefaultendpunct}{\mcitedefaultseppunct}\relax
\EndOfBibitem
\bibitem[Zhong \latin{et~al.}(2016)Zhong, Quhe, Wang, Ni, Ye, Song, Pan, Yang,
  Yang, Lei, Shi, and Lu]{Zhong2016}
Zhong,~H.; Quhe,~R.; Wang,~Y.; Ni,~Z.; Ye,~M.; Song,~Z.; Pan,~Y.; Yang,~J.;
  Yang,~L.; Lei,~M.; Shi,~J.; Lu,~J. {Interfacial Properties of Monolayer and
  Bilayer MoS$_2$ Contacts with Metals: Beyond the Energy Band Calculations}.
  \emph{Sci. Rep.} \textbf{2016}, \emph{6}, 21786\relax
\mciteBstWouldAddEndPuncttrue
\mciteSetBstMidEndSepPunct{\mcitedefaultmidpunct}
{\mcitedefaultendpunct}{\mcitedefaultseppunct}\relax
\EndOfBibitem
\bibitem[Choi \latin{et~al.}(2014)Choi, Shaolin, and Yang]{Choi2014}
Choi,~S.; Shaolin,~Z.; Yang,~W. {Layer-Number-Dependent Work Function of
  MoS$_2$ Nanoflakes}. \emph{J. Korean Phys. Soc.} \textbf{2014}, \emph{64},
  1550--1555\relax
\mciteBstWouldAddEndPuncttrue
\mciteSetBstMidEndSepPunct{\mcitedefaultmidpunct}
{\mcitedefaultendpunct}{\mcitedefaultseppunct}\relax
\EndOfBibitem
\bibitem[Zhao \latin{et~al.}(2018)Zhao, Ding, Chen, Dan, Poh, Fu, Pennycook,
  Zhou, and Loh]{Zhao2018}
Zhao,~X.; Ding,~Z.; Chen,~J.; Dan,~J.; Poh,~S.~M.; Fu,~W.; Pennycook,~S.~J.;
  Zhou,~W.; Loh,~K.~P. {Strain Modulation by van der Waals Coupling in Bilayer
  Transition Metal Dichalcogenide}. \emph{ACS Nano} \textbf{2018}, \emph{12},
  1940--1948\relax
\mciteBstWouldAddEndPuncttrue
\mciteSetBstMidEndSepPunct{\mcitedefaultmidpunct}
{\mcitedefaultendpunct}{\mcitedefaultseppunct}\relax
\EndOfBibitem
\bibitem[{Nalin Mehta} \latin{et~al.}(2020){Nalin Mehta}, Mo, Pourtois, Dabral,
  Groven, Bender, Favia, Caymax, and Vandervorst]{NalinMehta2020}
{Nalin Mehta},~A.; Mo,~J.; Pourtois,~G.; Dabral,~A.; Groven,~B.; Bender,~H.;
  Favia,~P.; Caymax,~M.; Vandervorst,~W. {Grain-Boundary-Induced Strain and
  Distortion in Epitaxial Bilayer MoS$_ 2$ Lattice}. \emph{J. Phys. Chem. C}
  \textbf{2020}, \emph{124}, 6472--6478\relax
\mciteBstWouldAddEndPuncttrue
\mciteSetBstMidEndSepPunct{\mcitedefaultmidpunct}
{\mcitedefaultendpunct}{\mcitedefaultseppunct}\relax
\EndOfBibitem
\bibitem[Peelaers and {Van de Walle}(2012)Peelaers, and {Van de
  Walle}]{Peelaers2012}
Peelaers,~H.; {Van de Walle},~C.~G. {Effects of Strain on Band Structure and
  Effective Masses in MoS$_2$}. \emph{Phys. Rev. B} \textbf{2012}, \emph{86},
  241401\relax
\mciteBstWouldAddEndPuncttrue
\mciteSetBstMidEndSepPunct{\mcitedefaultmidpunct}
{\mcitedefaultendpunct}{\mcitedefaultseppunct}\relax
\EndOfBibitem
\bibitem[Yu \latin{et~al.}(2015)Yu, Xiong, Eshun, Yuan, and Li]{Yu2015}
Yu,~S.; Xiong,~H.~D.; Eshun,~K.; Yuan,~H.; Li,~Q. {Phase Transition, Effective
  Mass and Carrier Mobility of MoS$_2$ Monolayer under Tensile Strain}.
  \emph{Appl. Surf. Sci.} \textbf{2015}, \emph{325}, 27--32\relax
\mciteBstWouldAddEndPuncttrue
\mciteSetBstMidEndSepPunct{\mcitedefaultmidpunct}
{\mcitedefaultendpunct}{\mcitedefaultseppunct}\relax
\EndOfBibitem
\bibitem[Ehlen \latin{et~al.}(2018)Ehlen, Hall, Senkovskiy, Hell, Li, Herman,
  Smirnov, Fedorov, {Yu Voroshnin}, {Di Santo}, Petaccia, Michely, and
  Gr{\"{u}}neis]{Ehlen2018}
Ehlen,~N.; Hall,~J.; Senkovskiy,~B.~V.; Hell,~M.; Li,~J.; Herman,~A.;
  Smirnov,~D.; Fedorov,~A.; {Yu Voroshnin},~V.; {Di Santo},~G.; Petaccia,~L.;
  Michely,~T.; Gr{\"{u}}neis,~A. {Narrow Photoluminescence and Raman Peaks of
  Epitaxial MoS$_2$ on Graphene/Ir(111)}. \emph{2D Mater.} \textbf{2018},
  \emph{6}, 011006\relax
\mciteBstWouldAddEndPuncttrue
\mciteSetBstMidEndSepPunct{\mcitedefaultmidpunct}
{\mcitedefaultendpunct}{\mcitedefaultseppunct}\relax
\EndOfBibitem
\bibitem[Pletikosi{\'{c}} \latin{et~al.}(2009)Pletikosi{\'{c}}, Kralj, Pervan,
  Brako, Coraux, N'Diaye, Busse, and Michely]{Pletikosic2009}
Pletikosi{\'{c}},~I.; Kralj,~M.; Pervan,~P.; Brako,~R.; Coraux,~J.;
  N'Diaye,~A.~T.; Busse,~C.; Michely,~T. {Dirac Cones and Minigaps for Graphene
  on Ir(111)}. \emph{Phys. Rev. Lett.} \textbf{2009}, \emph{102}, 056808\relax
\mciteBstWouldAddEndPuncttrue
\mciteSetBstMidEndSepPunct{\mcitedefaultmidpunct}
{\mcitedefaultendpunct}{\mcitedefaultseppunct}\relax
\EndOfBibitem
\bibitem[Laturia \latin{et~al.}(2018)Laturia, {Van de Put}, and
  Vandenberghe]{Laturia2018}
Laturia,~A.; {Van de Put},~M.~L.; Vandenberghe,~W.~G. {Dielectric Properties of
  Hexagonal Boron Nitride and Transition Metal Dichalcogenides: From Monolayer
  to Bulk}. \emph{npj 2D Mater. Appl.} \textbf{2018}, \emph{2}, 6\relax
\mciteBstWouldAddEndPuncttrue
\mciteSetBstMidEndSepPunct{\mcitedefaultmidpunct}
{\mcitedefaultendpunct}{\mcitedefaultseppunct}\relax
\EndOfBibitem
\bibitem[Qiu \latin{et~al.}(2018)Qiu, Zhao, Hu, Yue, Ren, and Yuan]{Qiu2018}
Qiu,~B.; Zhao,~X.; Hu,~G.; Yue,~W.; Ren,~J.; Yuan,~X. {Optical Properties of
  Graphene/MoS$_2$ Heterostructure: First Principles Calculations}.
  \emph{Nanomaterials} \textbf{2018}, \emph{8}, 962\relax
\mciteBstWouldAddEndPuncttrue
\mciteSetBstMidEndSepPunct{\mcitedefaultmidpunct}
{\mcitedefaultendpunct}{\mcitedefaultseppunct}\relax
\EndOfBibitem
\bibitem[Schumacher \latin{et~al.}(2013)Schumacher, Wehling, Lazi{\'{c}},
  Runte, F{\"{o}}rster, Busse, Petrovi{\'{c}}, Kralj, Bl{\"{u}}gel, Atodiresei,
  Caciuc, and Michely]{Schumacher2013NL}
Schumacher,~S.; Wehling,~T.~O.; Lazi{\'{c}},~P.; Runte,~S.;
  F{\"{o}}rster,~D.~F.; Busse,~C.; Petrovi{\'{c}},~M.; Kralj,~M.;
  Bl{\"{u}}gel,~S.; Atodiresei,~N.; Caciuc,~V.; Michely,~T. {The Backside of
  Graphene: Manipulating Adsorption by Intercalation}. \emph{Nano Lett.}
  \textbf{2013}, \emph{13}, 5013--5019\relax
\mciteBstWouldAddEndPuncttrue
\mciteSetBstMidEndSepPunct{\mcitedefaultmidpunct}
{\mcitedefaultendpunct}{\mcitedefaultseppunct}\relax
\EndOfBibitem
\bibitem[Schr{\"{o}}der \latin{et~al.}(2016)Schr{\"{o}}der, Petrovi{\'{c}},
  Gerber, Mart{\'{i}}nez-Galera, Gr{\aa}n{\"{a}}s, Arman, Herbig, Schnadt,
  Kralj, Knudsen, and Michely]{Schroder2016}
Schr{\"{o}}der,~U.~A.; Petrovi{\'{c}},~M.; Gerber,~T.;
  Mart{\'{i}}nez-Galera,~A.~J.; Gr{\aa}n{\"{a}}s,~E.; Arman,~M.~A.; Herbig,~C.;
  Schnadt,~J.; Kralj,~M.; Knudsen,~J.; Michely,~T. {Core Level Shifts of
  Intercalated Graphene}. \emph{2D Mater.} \textbf{2016}, \emph{4},
  015013\relax
\mciteBstWouldAddEndPuncttrue
\mciteSetBstMidEndSepPunct{\mcitedefaultmidpunct}
{\mcitedefaultendpunct}{\mcitedefaultseppunct}\relax
\EndOfBibitem
\bibitem[Pulkin and Yazyev(2016)Pulkin, and Yazyev]{Pulkin2016}
Pulkin,~A.; Yazyev,~O.~V. {Spin- and Valley-Polarized Transport across Line
  Defects in Monolayer MoS$_2$}. \emph{Phys. Rev. B} \textbf{2016}, \emph{93},
  041419\relax
\mciteBstWouldAddEndPuncttrue
\mciteSetBstMidEndSepPunct{\mcitedefaultmidpunct}
{\mcitedefaultendpunct}{\mcitedefaultseppunct}\relax
\EndOfBibitem
\bibitem[Park \latin{et~al.}(2019)Park, Xue, Mouis, Triozon, and
  Cresti]{Park2019}
Park,~J.; Xue,~K.-H.; Mouis,~M.; Triozon,~F.; Cresti,~A. {Electron Transport
  Properties of Mirror Twin Grain Boundaries in Molybdenum Disulfide: Impact of
  Disorder}. \emph{Phys. Rev. B} \textbf{2019}, \emph{100}, 235403\relax
\mciteBstWouldAddEndPuncttrue
\mciteSetBstMidEndSepPunct{\mcitedefaultmidpunct}
{\mcitedefaultendpunct}{\mcitedefaultseppunct}\relax
\EndOfBibitem
\bibitem[Ohtomo and Hwang(2004)Ohtomo, and Hwang]{Ohtomo2004}
Ohtomo,~A.; Hwang,~H.~Y. {A High-Mobility Electron Gas at the
  LaAlO$_3$/SrTiO$_3$ Heterointerface}. \emph{Nature} \textbf{2004},
  \emph{427}, 423--426\relax
\mciteBstWouldAddEndPuncttrue
\mciteSetBstMidEndSepPunct{\mcitedefaultmidpunct}
{\mcitedefaultendpunct}{\mcitedefaultseppunct}\relax
\EndOfBibitem
\bibitem[Gibertini \latin{et~al.}(2014)Gibertini, Pizzi, and
  Marzari]{Gibertini2014}
Gibertini,~M.; Pizzi,~G.; Marzari,~N. {Engineering Polar Discontinuities in
  Honeycomb Lattices}. \emph{Nat. Commun.} \textbf{2014}, \emph{5}, 5157\relax
\mciteBstWouldAddEndPuncttrue
\mciteSetBstMidEndSepPunct{\mcitedefaultmidpunct}
{\mcitedefaultendpunct}{\mcitedefaultseppunct}\relax
\EndOfBibitem
\bibitem[Martinez-Gordillo and Pruneda(2015)Martinez-Gordillo, and
  Pruneda]{Martinez-Gordillo2015}
Martinez-Gordillo,~R.; Pruneda,~M. {Polar Discontinuities and 1D Interfaces in
  Monolayered Materials}. \emph{Prog. Surf. Sci.} \textbf{2015}, \emph{90},
  444--463\relax
\mciteBstWouldAddEndPuncttrue
\mciteSetBstMidEndSepPunct{\mcitedefaultmidpunct}
{\mcitedefaultendpunct}{\mcitedefaultseppunct}\relax
\EndOfBibitem
\bibitem[Park \latin{et~al.}(2014)Park, Lee, Liu, Clark, Durand, Park, Sumpter,
  Baddorf, Mohsin, Yoon, Gu, and Li]{Park2014}
Park,~J.; Lee,~J.; Liu,~L.; Clark,~K.~W.; Durand,~C.; Park,~C.; Sumpter,~B.~G.;
  Baddorf,~A.~P.; Mohsin,~A.; Yoon,~M.; Gu,~G.; Li,~A.-P. {Spatially Resolved
  One-Dimensional Boundary States in Graphene-Hexagonal Boron Nitride Planar
  Heterostructures}. \emph{Nat. Commun.} \textbf{2014}, \emph{5}, 5403\relax
\mciteBstWouldAddEndPuncttrue
\mciteSetBstMidEndSepPunct{\mcitedefaultmidpunct}
{\mcitedefaultendpunct}{\mcitedefaultseppunct}\relax
\EndOfBibitem
\bibitem[Mannhart and Schlom(2010)Mannhart, and Schlom]{Mannhart2010a}
Mannhart,~J.; Schlom,~D.~G. {Oxide Interfaces - An Opportunity for
  Electronics}. \emph{Science} \textbf{2010}, \emph{327}, 1607--1611\relax
\mciteBstWouldAddEndPuncttrue
\mciteSetBstMidEndSepPunct{\mcitedefaultmidpunct}
{\mcitedefaultendpunct}{\mcitedefaultseppunct}\relax
\EndOfBibitem
\bibitem[Pai \latin{et~al.}(2018)Pai, Tylan-Tyler, Irvin, and Levy]{Pai2018}
Pai,~Y.-Y.; Tylan-Tyler,~A.; Irvin,~P.; Levy,~J. {Physics of SrTiO$_3$-Based
  Heterostructures and Nanostructures: A Review}. \emph{Reports Prog. Phys.}
  \textbf{2018}, \emph{81}, 036503\relax
\mciteBstWouldAddEndPuncttrue
\mciteSetBstMidEndSepPunct{\mcitedefaultmidpunct}
{\mcitedefaultendpunct}{\mcitedefaultseppunct}\relax
\EndOfBibitem
\bibitem[Feng \latin{et~al.}(2012)Feng, Qian, Huang, and Li]{Feng2012}
Feng,~J.; Qian,~X.; Huang,~C.~W.; Li,~J. {Strain-Engineered Artificial Atom as
  a Broad-Spectrum Solar Energy Funnel}. \emph{Nat. Photonics} \textbf{2012},
  \emph{6}, 866--872\relax
\mciteBstWouldAddEndPuncttrue
\mciteSetBstMidEndSepPunct{\mcitedefaultmidpunct}
{\mcitedefaultendpunct}{\mcitedefaultseppunct}\relax
\EndOfBibitem
\bibitem[Branny \latin{et~al.}(2017)Branny, Kumar, Proux, and
  Gerardot]{Branny2017}
Branny,~A.; Kumar,~S.; Proux,~R.; Gerardot,~B.~D. {Deterministic Strain-Induced
  Arrays of Quantum Emitters in a Two-Dimensional Semiconductor}. \emph{Nat.
  Commun.} \textbf{2017}, \emph{8}, 15053\relax
\mciteBstWouldAddEndPuncttrue
\mciteSetBstMidEndSepPunct{\mcitedefaultmidpunct}
{\mcitedefaultendpunct}{\mcitedefaultseppunct}\relax
\EndOfBibitem
\bibitem[Carozo \latin{et~al.}(2017)Carozo, Wang, Fujisawa, Carvalho, McCreary,
  Feng, Lin, Zhou, Perea-L{\'{o}}pez, El{\'{i}}as, Kabius, Crespi, and
  Terrones]{Carozo2017}
Carozo,~V.; Wang,~Y.; Fujisawa,~K.; Carvalho,~B.~R.; McCreary,~A.; Feng,~S.;
  Lin,~Z.; Zhou,~C.; Perea-L{\'{o}}pez,~N.; El{\'{i}}as,~A.~L.; Kabius,~B.;
  Crespi,~V.~H.; Terrones,~M. {Optical Identification of Sulfur Vacancies:
  Bound Excitons at the Edges of Monolayer Tungsten Disulfide}. \emph{Sci.
  Adv.} \textbf{2017}, \emph{3}, e1602813\relax
\mciteBstWouldAddEndPuncttrue
\mciteSetBstMidEndSepPunct{\mcitedefaultmidpunct}
{\mcitedefaultendpunct}{\mcitedefaultseppunct}\relax
\EndOfBibitem
\bibitem[Dubey \latin{et~al.}(2017)Dubey, Lisi, Nayak, Herziger, Nguyen, {Le
  Quang}, Cherkez, Gonz{\'{a}}lez, Dappe, Watanabe, Taniguchi, Magaud, Mallet,
  Veuillen, Arenal, Marty, Renard, Bendiab, Coraux, and Bouchiat]{Dubey2017}
Dubey,~S.; Lisi,~S.; Nayak,~G.; Herziger,~F.; Nguyen,~V.~D.; {Le Quang},~T.;
  Cherkez,~V.; Gonz{\'{a}}lez,~C.; Dappe,~Y.~J.; Watanabe,~K.; Taniguchi,~T.;
  Magaud,~L.; Mallet,~P.; Veuillen,~J.~Y.; Arenal,~R.; Marty,~L.; Renard,~J.;
  Bendiab,~N.; Coraux,~J.; Bouchiat,~V. {Weakly Trapped, Charged, and Free
  Excitons in Single-Layer MoS$_2$ in the Presence of Defects, Strain, and
  Charged Impurities}. \emph{ACS Nano} \textbf{2017}, \emph{11},
  11206--11216\relax
\mciteBstWouldAddEndPuncttrue
\mciteSetBstMidEndSepPunct{\mcitedefaultmidpunct}
{\mcitedefaultendpunct}{\mcitedefaultseppunct}\relax
\EndOfBibitem
\bibitem[Zheng \latin{et~al.}(2019)Zheng, Chen, Huang, Gogoi, Li, Li,
  Trevisanutto, Wang, Pennycook, Wee, and Quek]{Zheng2019}
Zheng,~Y.~J.; Chen,~Y.; Huang,~Y.~L.; Gogoi,~P.~K.; Li,~M.~Y.; Li,~L.~J.;
  Trevisanutto,~P.~E.; Wang,~Q.; Pennycook,~S.~J.; Wee,~A.~T.; Quek,~S.~Y.
  {Point Defects and Localized Excitons in 2D WSe$_2$}. \emph{ACS Nano}
  \textbf{2019}, \emph{13}, 6050--6059\relax
\mciteBstWouldAddEndPuncttrue
\mciteSetBstMidEndSepPunct{\mcitedefaultmidpunct}
{\mcitedefaultendpunct}{\mcitedefaultseppunct}\relax
\EndOfBibitem
\bibitem[van Gastel \latin{et~al.}(2009)van Gastel, N'Diaye, Wall, Coraux,
  Busse, Buckanie, {Meyer zu Heringdorf}, {Horn von Hoegen}, Michely, and
  Poelsema]{VanGastel2009}
van Gastel,~R.; N'Diaye,~A.~T.; Wall,~D.; Coraux,~J.; Busse,~C.;
  Buckanie,~N.~M.; {Meyer zu Heringdorf},~F.-J.; {Horn von Hoegen},~M.;
  Michely,~T.; Poelsema,~B. {Selecting a Single Orientation for Millimeter
  Sized Graphene Sheets}. \emph{Appl. Phys. Lett.} \textbf{2009}, \emph{95},
  121901\relax
\mciteBstWouldAddEndPuncttrue
\mciteSetBstMidEndSepPunct{\mcitedefaultmidpunct}
{\mcitedefaultendpunct}{\mcitedefaultseppunct}\relax
\EndOfBibitem
\bibitem[Morgenstern(2003)]{Morgenstern2003}
Morgenstern,~M. {Probing the Local Density of States of Dilute Electron Systems
  in Different Dimensions}. \emph{Surf. Rev. Lett.} \textbf{2003}, \emph{10},
  933--962\relax
\mciteBstWouldAddEndPuncttrue
\mciteSetBstMidEndSepPunct{\mcitedefaultmidpunct}
{\mcitedefaultendpunct}{\mcitedefaultseppunct}\relax
\EndOfBibitem
\bibitem[Kresse and Hafner(1993)Kresse, and Hafner]{Kresse1993}
Kresse,~G.; Hafner,~J. {\textit{Ab Initio} Molecular Dynamics for Open-Shell
  Transition Metals}. \emph{Phys. Rev. B} \textbf{1993}, \emph{48},
  13115--13118\relax
\mciteBstWouldAddEndPuncttrue
\mciteSetBstMidEndSepPunct{\mcitedefaultmidpunct}
{\mcitedefaultendpunct}{\mcitedefaultseppunct}\relax
\EndOfBibitem
\bibitem[Kresse and Furthm{\"{u}}ller(1996)Kresse, and
  Furthm{\"{u}}ller]{Kresse1996}
Kresse,~G.; Furthm{\"{u}}ller,~J. {Efficiency of \textit{Ab-Initio} Total
  Energy Calculations for Metals and Semiconductors Using a Plane-Wave Basis
  Set}. \emph{Comput. Mater. Sci.} \textbf{1996}, \emph{6}, 15--50\relax
\mciteBstWouldAddEndPuncttrue
\mciteSetBstMidEndSepPunct{\mcitedefaultmidpunct}
{\mcitedefaultendpunct}{\mcitedefaultseppunct}\relax
\EndOfBibitem
\bibitem[Perdew \latin{et~al.}(1996)Perdew, Burke, and Ernzerhof]{Perdew1996}
Perdew,~J.~P.; Burke,~K.; Ernzerhof,~M. {Generalized Gradient Approximation
  Made Simple}. \emph{Phys. Rev. Lett.} \textbf{1996}, \emph{77},
  3865--3868\relax
\mciteBstWouldAddEndPuncttrue
\mciteSetBstMidEndSepPunct{\mcitedefaultmidpunct}
{\mcitedefaultendpunct}{\mcitedefaultseppunct}\relax
\EndOfBibitem
\bibitem[Grimme(2006)]{Grimme2006}
Grimme,~S. {Semiempirical GGA-Type Density Functional Constructed with a
  Long-Range Dispersion Correction}. \emph{J. Comput. Chem.} \textbf{2006},
  \emph{27}, 1787--1799\relax
\mciteBstWouldAddEndPuncttrue
\mciteSetBstMidEndSepPunct{\mcitedefaultmidpunct}
{\mcitedefaultendpunct}{\mcitedefaultseppunct}\relax
\EndOfBibitem
\end{mcitethebibliography}

\end{document}